\documentclass[10pt,twocolumn,aps,prb,balancelastpage,longbibliography]{revtex4-1}
\usepackage[latin9]{inputenc}
\usepackage{color}
\usepackage{verbatim}
\usepackage{amsmath}
\usepackage{amssymb,ulem}
\usepackage{graphicx,cellspace}
\usepackage{esint}
\usepackage[unicode=true,pdfusetitle,
 bookmarks=true,bookmarksnumbered=false,bookmarksopen=false,
 breaklinks=false,pdfborder={0 0 1},backref=false,colorlinks=true]{hyperref}
\usepackage{breakurl}
\usepackage{epstopdf}
\usepackage{feynmp-auto}
\usepackage{youngtab}
\usepackage{ytableau}

\makeatletter

\@ifundefined{textcolor}{}
{%
 \definecolor{BLACK}{gray}{0}
 \definecolor{WHITE}{gray}{1}
 \definecolor{RED}{rgb}{1,0,0}
 \definecolor{GREEN}{rgb}{0,1,0}
 \definecolor{BLUE}{rgb}{0,0,1}
 \definecolor{CYAN}{cmyk}{1,0,0,0}
 \definecolor{MAGENTA}{cmyk}{0,1,0,0}
 \definecolor{YELLOW}{cmyk}{0,0,1,0}
}

\usepackage{subfigure}\usepackage{bm}

\newcommand{\braket}[2]{\langle #1 | #2 \rangle}
\newcommand{\tr}{\mbox{tr}}

\newcommand{\bv}[1]{\mathbf{#1}}

\makeatother

\begin{document}

\title{Strongly Interacting Phases of Metallic Wires in Strong Magnetic Field}

\author{Daniel Bulmash}

\affiliation{Department of Physics, Stanford University, Stanford, California 94305-4045, USA}

\author{Chao-Ming Jian}

\affiliation{Kavli Institute for Theoretical Physics, University of California, Santa Barbara, California, 93106, USA}
\affiliation{Station Q, Microsoft Research, Santa Barbara, California 93106-6105, USA}

\author{Xiao-Liang Qi}

\affiliation{Department of Physics, Stanford University, Stanford, California 94305-4045, USA}

\date{\today}
\begin{abstract}
We investigate theoretically an interacting metallic wire with a strong magnetic field directed along its length and show that it is a new and highly tunable one-dimensional system. By considering a suitable change in spatial geometry, we map the problem in the zeroth Landau level with Landau level degeneracy $N$ to one-dimensional fermions with an $N$-component pseudospin degree of freedom and $SU(2)$-symmetric interactions. This mapping allows us to establish the phase diagram as a function of the interactions for small $N$ (and make conjectures for large $N$) using renormalization group and bosonization techniques. We find pseudospin-charge separation with a gapless $U(1)$ charge sector and several possible strong-coupling phases in the pseudospin sector. For odd $N$, we find a fluctuating pseudospin-singlet charge density wave phase and a fluctuating pseudospin-singlet superconducting phase which are topologically distinct. For even $N>2$, similar phases exist, although they are not topologically distinct, and an additional, novel pseudospin-gapless phase appears. We discuss experimental conditions for observing our proposals.

\end{abstract}
\maketitle

\section{Introduction}
\label{sec:intro}

Interacting quantum systems in one spatial dimension exhibit many exotic behaviors, such as Luttinger liquid phases and other phases with quasi-long-range order\cite{Voit1DFermiLiquids}. Remarkably, these behaviors are often tractable theoretically thanks to powerful tools special to one dimension (1D), such as bosonization\cite{BosonizationBook} and 1+1D conformal field theory (CFT) techniques\cite{CFTBook}. There are a wide range of systems which can be treated with such tools, including spin chains\cite{AffleckHaldane}, 1D metals\cite{Voit1DFermiLiquids}, and coupled wires\cite{KivelsonCoupledWires,FradkinKivelsonWires,KaneFQHWires}, but the underlying degrees of freedom in the 1D problem are typically not possible to tune, in the sense that spin chains are always (after fermionization) built from a fixed number of colors of spin-1/2 fermions and 1D metals are always built from spin-1/2 fermions. 

In this paper, we consider a spinless, interacting metallic wire with strong magnetic field directed along its length and relate it to a new class of 1D systems: interacting metals whose electrons have a large (pseudo)spin. This is particularly interesting because the fact that the magnetic field changes the Landau level degeneracy in the first problem will map onto a tunable number of (degenerate) spin states in the second problem. 

For the simplest intuition about how to treat the problem of the wire in field, consider semiclassical electrons traveling in three dimensions in a magnetic field $B$. They move freely along the direction of the field, but in the plane perpendicular to the field, they move in cyclotron orbits whose radius goes as $1/B$. At strong field, the motion thus becomes increasingly one-dimensional, similar to the plasma physics concept of magnetic confinement, and the number of non-overlapping orbits that fit into a wire scales as $B$. In more quantum language, consider a metal in a magnetic field strong enough that only the zeroth Landau level (ZLL) is occupied at every momentum along the field. Kinetic energy is quenched in directions perpendicular to the field, so naively the degenerate Landau level states are like one-dimensional wires which are coupled only by electronic interactions, and the degeneracy scales with $B$.

However, in the quantum case there is a key difference between the ZLL problem and coupled wires. As a consequence of the nontrivial topological invariant of the Landau level\cite{TKNN}, no orthogonal basis for the ZLL can have wavefunctions which are local in both directions perpendicular to the field. Since electron-electron interactions are local in real space, this means that there is no natural choice of basis in which the interaction between basis states is local. Another problem is that the choice of basis makes magnetic translation symmetry implicit, making it difficult to make approximations while preserving the symmetry.

Motivated by the problems of the coupled wire picture, in this paper we propose an alternative approach to this problem which explicitly preserves symmetry. We map a metallic wire in the quantum limit with an $N$-fold degenerate ZLL to a large-pseudospin one-dimensional wire with $N$ degenerate spin states. Magnetic translation symmetry is mapped to an $SU(2)$ symmetry of the pseudospin. (The boundary of the wire, which breaks magnetic translation symmetry, is mapped to an $SU(2)$-breaking external field.) Although this mapping is a small modification of one already known\cite{JainBook} at the level of non-interacting electrons, our main insight is that the resulting one-dimensionality and symmetry make the interacting problem tractable. We are able to apply the powerful machinery of both Abelian and non-Abelian bosonization, along with conformal field theory techniques, to elucidate the phase diagram as a function of generic interaction parameters.

There has been considerable previous work on interacting bulk metals in the zeroth Landau level. On the theory side, many approaches of varying sophistication have been used, resulting in predictions of density waves\cite{CelliSDWStrongField,FukuyamaCDWField}, exciton insulators\cite{AbrikosovMetalStrongField}, superconductors\cite{RasoltSCInHighField} (SC), and marginal Fermi liquids\cite{Yakovenko}. Experimentally, there is evidence for field-induced transitions to an insulating state in bulk bismuth\cite{MiuraBiSemiconducting,HirumaFieldBi} and graphite\cite{IyeGraphite}, which have been understood as charge density wave (CDW) transitions\cite{YoshiokaGraphiteCDW} but are still being studied. In contrast, our interest is in using a wire geometry in order to more clearly bring out the quasi-one-dimensionality induced by the magnetic field, and to more easily apply 1D tools.

A major technical strength of our approach is that the mapping to pseudospins accounts for interactions with range longer than the magnetic length, in contrast to previous work and any naive coupled-wire treatment.

Before proceeding, we summarize our phase diagram, which depends strongly on the parity of the Landau level degeneracy $N$. For odd $N$, we have identified three phases. One is a Luttinger liquid, having a gapless charge sector and a free pseudospin sector. The other two have a gapless charge sector and fully gapped pseudospin sector, and we argue that they are separated by a first-order transition. One has power-law correlations of the CDW order parameter and the other has power-law correlations of $p$-wave SC order; these phases are unusual because the power is tuned by $N$ (that is, by the magnetic field). For even $N>2$, we have identified four different phases, all of which have a gapless charge sector. One is again a Luttinger liquid. Two have a fully gapped pseudospin sector, with either power-law correlations of CDW order or s-wave SC order, and the transition between them can be second-order. Again the power laws can be tuned by $N$. The final phase is, to our knowledge, new: it has a gapless pseudospin sector, and we provide evidence that it has coexisting power-law correlations of pseudospin-density wave order and $p$-wave, pseudospin-triplet SC order.

The structure of this paper is as follows. In Section \ref{sec:noninteracting}, we discuss the non-interacting part of the model and construct the analogy between fermions in a wire and fermions on the spatial manifold $\mathbb{R} \times S^2$. In Section \ref{sec:interactions}, we write down the interacting Hamiltonian and cast it into a convenient form which makes its symmetry explicit. Sections \ref{sec:smallN} through \ref{sec:generalN} contain our main results. In Section \ref{sec:smallN}, for small $N$, we explicitly analyze our model through a perturbative renormalization group (RG) procedure and establish a phase diagram using non-Abelian bosonizaton. We identify the nature of the phases more explicitly using Abelian bosonization in Section \ref{sec:phaseID}. In Section \ref{sec:generalN}, we generalize the results of the previous two sections to conjectures about the phase diagram for all $N$. In Section \ref{sec:symmbreak}, we discuss the effect of symmetry-breaking perturbations in order to bring our results to bear on the experimentally relevant geometry. Section \ref{sec:largeN} relates our results to previously known ones in the bulk (large-$N$) limit. Finally, Section \ref{sec:discussion} consists of prospects for experimentally realizing these phases, open questions, and further discussion.

\section{Non-Interacting Model}
\label{sec:noninteracting}

In this section, we review the Landau level problem of spinless fermions on the wire $\mathbb{R} \times D^2$, where $D^2$ is the two-dimensional disk of radius $R$, and on the manifold $\mathbb{R} \times S^2$. We will build a connection between the two problems, and review the mapping from the lowest Landau level of the latter onto itinerant spinful 1D electrons. We will then use the latter model as the basis for much of the rest of the paper.

To establish conventions, we call the direction along the length of the wire $x$. The geometries are pictured in Table \ref{table:geometry}, along with a summary of the results of this section.

\begin{table*}
	\begin{tabular}{|l|l|Sl|} 
			\hline
			& \raisebox{0.6\height}{\includegraphics[width=5cm]{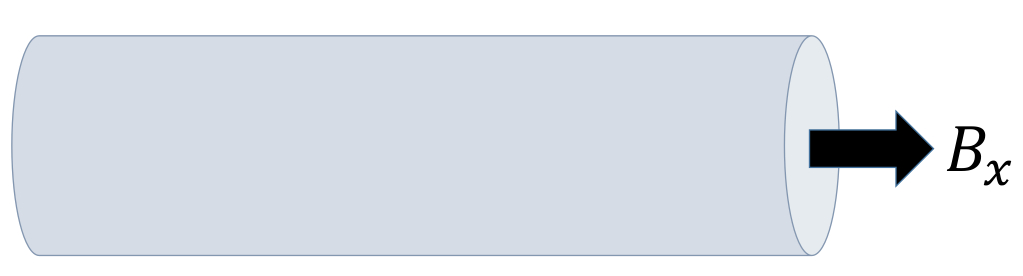}} & \includegraphics[width=5cm]{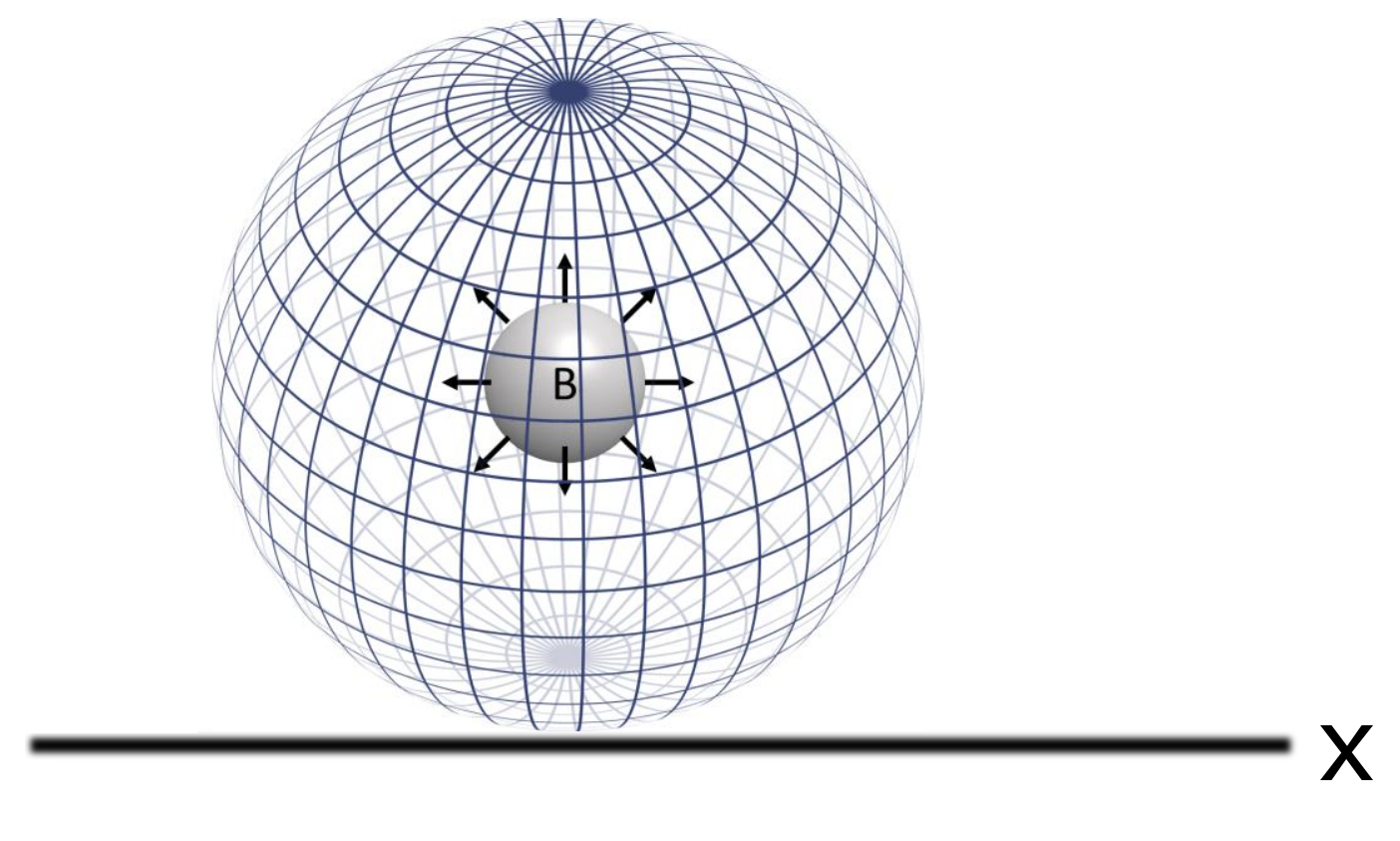}\\ \hline
			\textbf{Spatial geometry} & $\mathbb{R} \times D^2$ & $\mathbb{R} \times S^2$\\ \hline
			\textbf{Magnetic field} & Along $\hat{x}$ &  Monopole inside every $S^2$\\ \hline
			\textbf{ZLL degeneracy $N \equiv 2S_0+1$} & $\propto B$ & $\propto B$ \\ \hline
			\textbf{Quantum numbers in ZLL} & $k_x$, $L_x$ & $k_x$, $L_3$ \\ \hline
			\textbf{Magnetic translation symmetry} & Broken to $O(2)$ & $SU(2)$ \\ \hline
			
	\end{tabular}
	\caption{Comparison of a wire with a disk cross section ($\mathbb{R} \times D^2$ spatial geometry) and a wire with a spherical cross section ($\mathbb{R} \times S^2$ spatial geometry).}
		\label{table:geometry}
\end{table*}

\subsection{Landau Levels on the Disk and Sphere}

\begin{figure}
\subfigure[ ]{
\includegraphics[width=5cm]{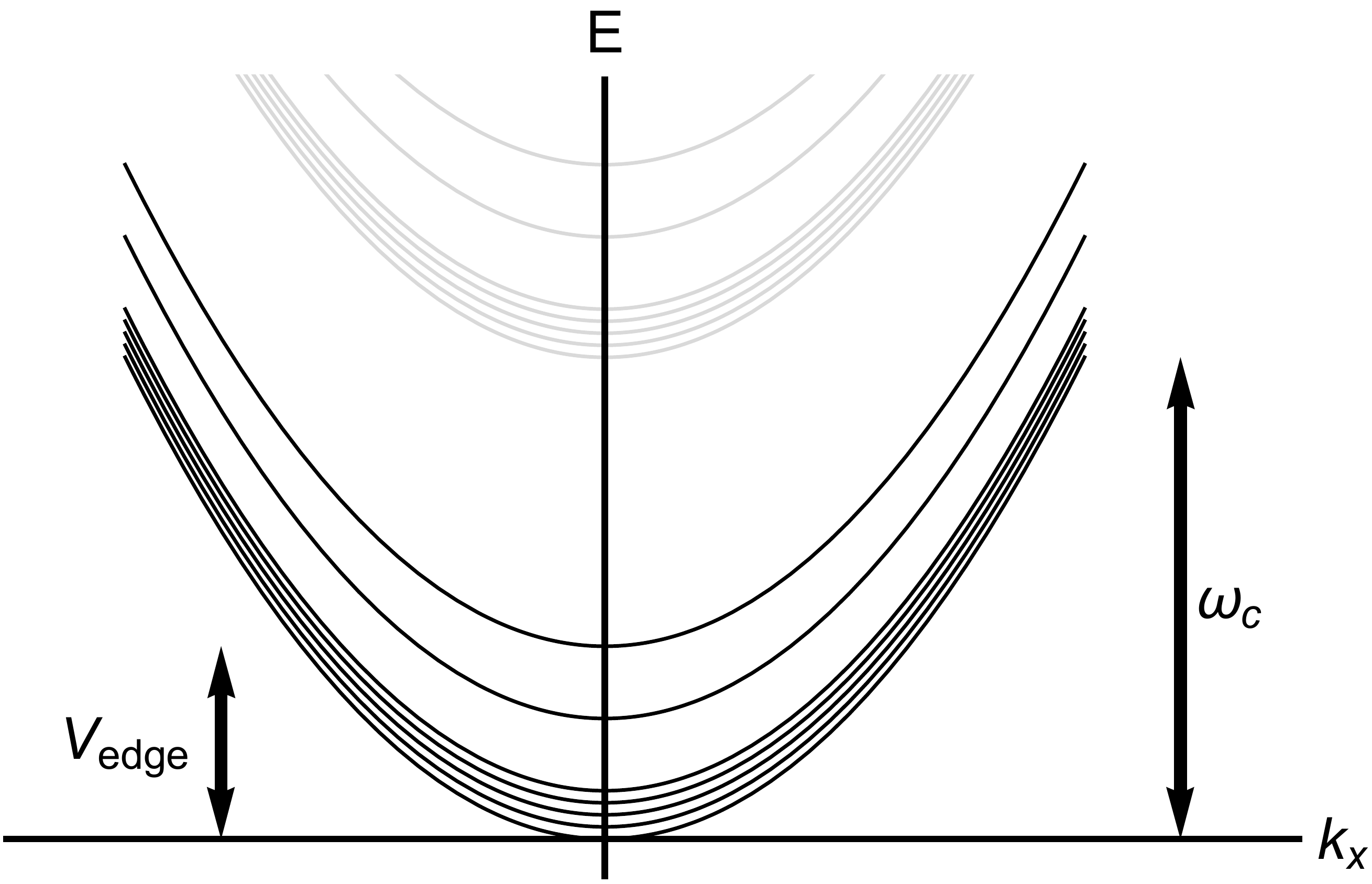}
\label{fig:brokenSymmDispersion}
}
\subfigure[ ]{
\includegraphics[width=5cm]{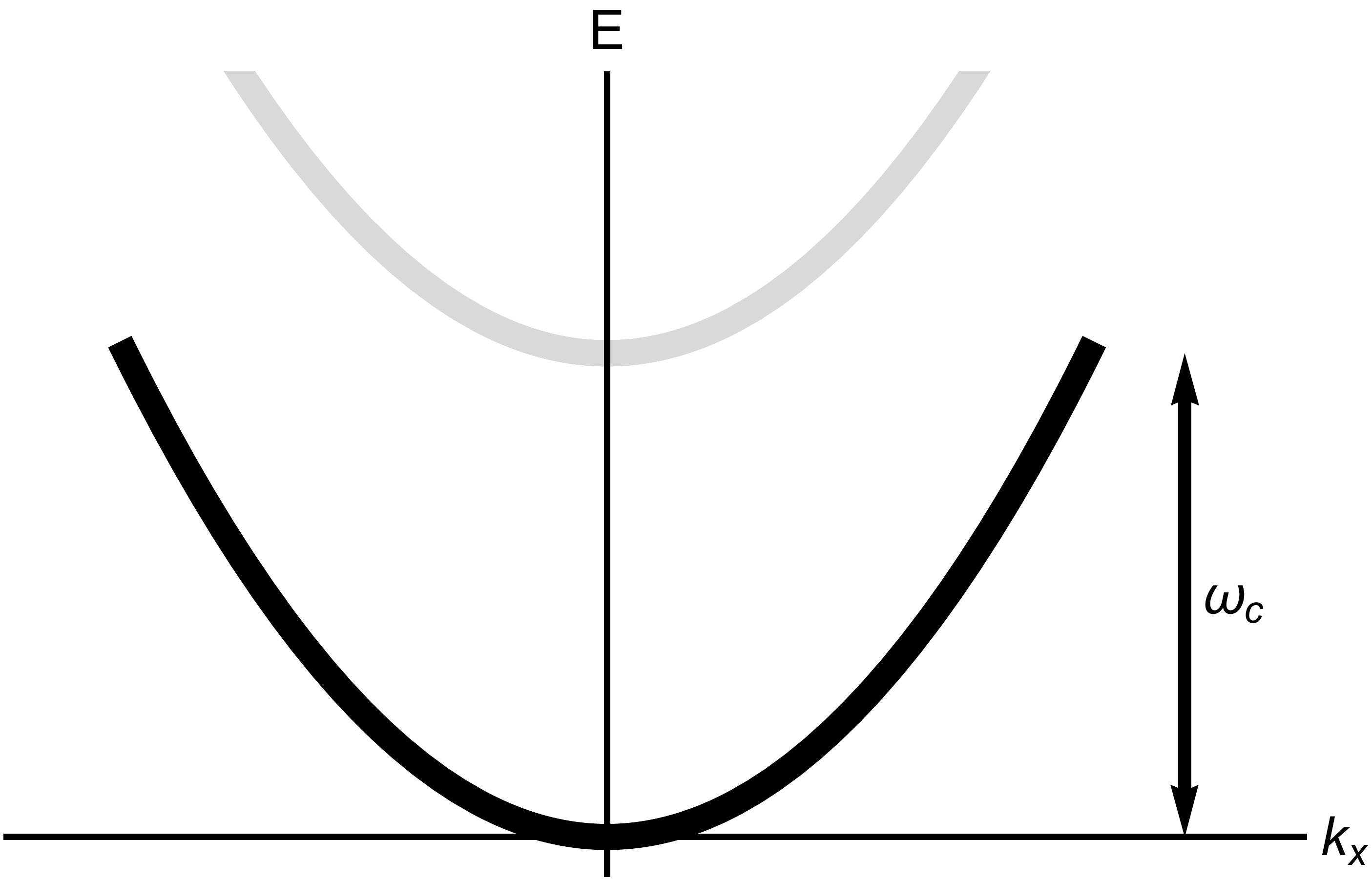}
\label{fig:symmDispersion}
}
\caption{(a) Energy spectrum of noninteracting electrons in a wire ($\mathbb{R} \times D^2$) geometry and strong magnetic field. (b) Energy spectrum of noninteracting electrons in an $\mathbb{R} \times S^2$ geometry and strong magnetic field. The dark curve is $(n_{\phi}+1)$-fold degenerate and the light one is $(n_{\phi}+3)$-fold degenerate. In both cases, the dark levels are in the $n=0$ Landau level and the light ones are in $n=1$.}
\end{figure}

We start by considering Schrodinger particles in a strong magnetic field along the $x$ direction, i.e. with Hamiltonian
\begin{equation}
H = \frac{(\bv{p}-e\bv{A})^2}{2m^{\ast}}
\end{equation}

where $m^{\ast}$ is the effective mass and $\bv{A}$ is the electromagnetic vector potential. We can always choose a gauge such that the eigenvalue $k_x$ of $p_x$ is a good quantum number. In the limit $R \rightarrow \infty$, this problem is simple; the spectrum forms Landau levels of energy
\begin{equation}
E_n(k_x) = \omega_c(n+1/2) + \frac{k_x^2}{2m^{\ast}}
\end{equation}
where $n$ is a non-negative integer and $\omega_c = eB/m^{\ast}$ is the cyclotron frequency. At fixed $k_x$, each Landau level has degeneracy approximately equal to the number of flux quanta $n_{\phi}$ penetrating a fixed-$x$ cross-section of the system. Working in symmetric gauge, as appropriate for the $\mathbb{R} \times D^2$ geometry, these degenerate states are localized in the radial direction and labeled by the integer eigenvalue $m$ of the angular momentum operator $L_x$. In the zeroth Landau level, the states have a spatial width of order $l_B = \sqrt{1/eB}$. At finite $R$, the degeneracy is broken due to the presence of the potential $V_{edge}$ associated with the boundary; those states which are radially localized close to the boundary have higher energy. The spectrum is shown schematically in Fig. \ref{fig:brokenSymmDispersion}.

This broken degeneracy arises from the boundary-induced loss of magnetic translation symmetry in the radial direction. The remaining symmetries are translations along $x$ and an $O(2)$ rotation symmetry. We would like more symmetry in order to better constrain the interacting problem. The reason, as discussed in the introduction, is that the nontrivial topological invariant\cite{TKNN} of a Landau level makes it impossible to form an orthogonal basis for the ZLL with wavefunctions local in both directions perpendicular to $x$. Therefore, interactions, projected to the ZLL, cannot be well-constrained by locality in any basis; with no locality and not much symmetry, there is no reason to expect the interacting problem to be tractable.

In order to enrich the symmetry, we change the spatial manifold to $\mathbb{R} \times S^2$. In this case, the wire has the spherical version of magnetic translation symmetry, which is an $SU(2)$ rotation symmetry. To see this, consider now Schrodinger electrons on a wire with a spherical cross-section, and suppose that every cross-section has a uniform, fixed flux piercing it. This requires a monopole inside the sphere, so the flux will be quantized to $n_{\phi} \in \mathbb{Z}$ flux quanta. The Hamiltonian is
\begin{equation}
H = \frac{\bv{\Lambda}^2}{2m^{\ast}R^2} + \frac{p_x^2}{2m^{\ast}}
\end{equation}
where $\bv{\Lambda} = \bv{r}\times \left( \bv{p}-e\bv{A}\right)$ is the canonical momentum on the sphere and $\bv{A}$ is a monopole vector potential. The radial component of $\bv{r}$ is \textit{not} related to $x$; it arises because writing $\bv{\Lambda}$ in this form requires embedding the $S^2$ in a fictitious extra spatial dimension. If the wire had finite length, then this geometry would indeed be analogous to a solid ball with a monopole placed in the center; the long direction of the wire would correspond to the radial direction on the ball. In this picture, though, the infinite radius limit would correspond to a semi-infinite wire, where $r=0$ corresponds to the single end of the wire, so this analogy is somewhat limited in the case we are considering.

Again, $p_x$ commutes with $H$, so we fix its eigenvalue $k_x$ to reduce to the Landau level problem in a spherical geometry. We briefly review standard facts about this problem\cite{JainBook}. The operator $\bv{L} = \bv{\Lambda} + n_{\phi}\hat{\bv{r}}/2$ commutes with the Hamiltonian and obeys the angular momentum algebra $[L_i,L_j] = i \varepsilon_{ijk}L_k$, where $i,j,k$ run over the three dimensions in which the $S^2$ is embedded and $\varepsilon$ is the Levi-Civita symbol. The good quantum numbers in the problem are the eigenvalues $k_x$, $l(l+1)$, and $m$ of the operators $p_x, \bv{L}^2$, and $L_3$ respectively, with $m=-l, -l+1, ... ,l$. Single-valuedness of the wavefunction only requires $2m-n_{\phi}$ to be an integer; hence $m$ can be a half-integer if $n_{\phi}$ is odd. The energy spectrum, shown in Fig. \ref{fig:symmDispersion}, is
\begin{equation}
E(l,m,k_z) = \frac{l(l+1)-(n_{\phi}/2)^2}{n_{\phi}}\omega_c + \frac{k_z^2}{2m^{\ast}}
\end{equation}
where $\omega_c = eB/m^{\ast}$ is the cyclotron frequency. There is also a restriction $l(l+1) \geq (n_{\phi}/2)^2$; therefore the lowest Landau level has $l=n_{\phi}/2$ and has degeneracy $N = n_{\phi}+1$.

Given that the angular momentum quantum numbers can be half-integers, the symmetry group corresponding to rotations of the spherical cross-section of the wire is $SU(2)$. Projecting to the lowest Landau level reduces all of the degrees of freedom on the $S^2$ to $N$ degenerate levels which transform as a pseudospin-$S_0$ representation of the $SU(2)$ symmetry, where 
\begin{equation}
S_0 = \frac{N-1}{2}
\end{equation}
This projected problem is therefore equivalent to purely one-dimensional itinerant fermions with a (possibly very large) pseudospin.

We expect that for large $N$, the lowest Landau level of the sphere problem and the disk (wire) problem should behave very similarly. In both cases, there is free propagation along the wire, and the finite-size directions are characterized by a large Landau level degeneracy. On the disk, at every $k_x$, all the states which far from the edge of the disk are nearly degenerate. The presence of the boundary breaks this degeneracy, but that effect is only strong near the edge. In the spherical case, the way to lift the Landau level degeneracy is by breaking $SU(2)$ symmetry.

The main idea of this paper is therefore to exploit the $SU(2)$ symmetry to understand the $\mathbb{R} \times S^2$ problem, and then add $SU(2)$-breaking perturbations to understand the physics of a wire.

\subsection{Low-Energy Non-Interacting Theory}

The rest of this paper will be devoted to finding instabilities of the non-interacting theory to interactions that are much weaker than the Landau level splitting and the bandwidth in $k_x$. To do the analysis, we need only consider the low-energy part of the non-interacting theory in the $\mathbb{R} \times S^2$ geometry, obtained by linearizing the dispersion of Fig. \ref{fig:symmDispersion} about the Fermi level. Define left- and right-moving fermions in the standard way
\begin{align}
\psi_m(x) &\sim \sum_{\pm} \int_{-\Lambda}^{\Lambda} \frac{dk}{2\pi} e^{i(k\pm k_F)x}\psi_m(k\pm k_F)\\
&\equiv e^{ik_Fx}\psi_{m,R}(x) + e^{-ik_Fx}\psi_{m,L}(x)
\end{align}
where $\Lambda \ll k_F$ is a momentum cutoff.

The low-energy Hamiltonian is then
\begin{equation}
H_0 =  \sum_{m=-S_0}^{S_0} \int dx \text{ } i v_F \left(\psi^{\dagger}_{m,L} \partial_x \psi_{m,L} - \psi^{\dagger}_{m,R}\partial_x \psi_{m,R}\right)
\label{eqn:H0}
\end{equation}
where $v_F$ is the Fermi velocity, which we set to 1. This Hamiltonian has an enormous $U(N)\otimes U(N)$ symmetry; the left- and right-movers may be transformed separately at the level of the low-energy theory. Interactions will break this symmetry to the nonchiral $SU(2)$ magnetic translation symmetry.

\subsection{Schrodinger vs. Weyl} 

In order to reach the zeroth Landau level, the carrier density needs to be low. In a standard metal or semiconductor, zero carrier density means the system is an insulator, and the above physics is not an appropriate description. In (type-I) Weyl semimetals\cite{Murakami2007,PyrochloreWeyl,WeylMultiLayer}, the Landau level at the Fermi energy still disperses linearly even at zero density. Such materials may be a promising system for realizing our proposal. To evaluate their suitability, we briefly compare and contrast Schrodinger and Weyl fermions as they pertain to our construction.

In either geometry, Schrodinger and Weyl fermions look very similar at low energies. The dispersion along $z$ is linear, and there is Landau level degeneracy; the Landau levels either have $SU(2)$ symmetry in the spherical case or magnetic translation symmetry in the bulk of the wire. There are three main differences. First, in the spherical case, the Landau level degeneracy $N$ for a given $n_{\phi}$ is $n_{\phi}+1$ for Schrodinger fermions and $n_{\phi}$ for Weyl fermions. Second, at fixed electron number $k_F$ is strongly dependent on the magnetic field in the Schrodinger case (since the Landau level degeneracy changes with field) but is set primarily by the Weyl point splitting in the Weyl case, with weak field-dependent corrections at finite doping above the Weyl points. Finally, the Landau level spacing is slightly different (at small momentum, it is proportional to $B$ for Schrodinger electrons and $\sqrt{B}$ for Weyl electrons).

These differences are inessential for the rest of our analysis; we abstract them away by fixing $N$ and $k_F$. Of course, these differences matter in a real experiment, as the difficulty of reaching the quantum limit with a given $N$ will depend on such factors; we will discuss this further in Section \ref{sec:discussion}.

For the rest of this paper, we use the $\mathbb{R} \times S^2$ geometry. We assume $SU(2)$ symmetry until section \ref{sec:symmbreak}, when we will make more contact with the wire geometry by investigating $SU(2)$-breaking perturbations.

\section{Structure of the Interactions}
\label{sec:interactions}

Starting from the free fermion fixed point, we now wish to write down the most relevant (in the RG sense) symmetry-respecting interaction terms. Four-fermion contact interactions are marginal at tree level; all other momentum-conserving interactions are irrelevant. Moreover, in the absence of fine-tuning to $k_F = \pi$, Umklapp scattering is forbidden. Finally, the interactions that we care about are non-chiral ones; fully chiral terms are exactly marginal and only renormalize velocities. As such the most relevant operators are left-right products of fermion bilinears, i.e. $\psi_{L,m}^{\dagger}A_{mm'}\psi_{L,m'}\psi_{R,n}^{\dagger}B_{nn'}\psi_{R,n'}$ where $A$ and $B$ are Hermitian $N \times N$ matrices. We now need to constrain $A$ and $B$ by symmetry.

\begin{figure}
\begin{fmffile}{RRLLInteraction}
\begin{fmfgraph*}(120,55)
	\fmfpen{thick}
	\fmfleft{i1,i2} 
	\fmfright{o1,o2}
	\fmf{fermion}{i1,v1,i2}
	\fmf{fermion}{o1,v2,o2}
	\fmf{photon,label=\small{$S;p$}}{v1,v2}
	\fmflabel{\small{$S_0;m;R$}}{i1}
	\fmflabel{\small{$S_0;m';R$}}{i2}
	\fmflabel{\small{$S_0;n;L$}}{o1}
	\fmflabel{\small{$S_0;n';L$}}{o2}
\end{fmfgraph*}
\end{fmffile}
\caption{Nonchiral interactions which are marginal at the free fermion fixed point. The labels $S_0$ and $m,m',n,n'$ indicate the $\bv{L}^2$ and the $L_3$ eigenvalue, respectively. The interaction is decomposed according to the angular momentum transfer $(L^2,L_3) = (S,p)$ from the left-mover to the right-mover.}
\label{fig:feynman}
\end{figure}
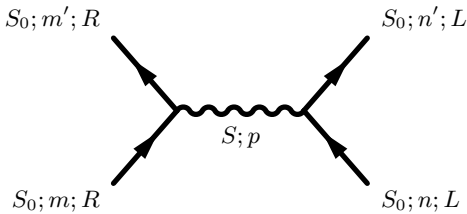

The interactions we want are shown in Fig. \ref{fig:feynman}. The interaction can be decomposed according to the angular momentum $(S,p)$ transferred from the left-mover to the right mover, where $S(S+1)$ and $p$ are the eigenvalues of of $\bv{L}^2$ and $L_3$ respectively. Here $S$ can range from $0$ to $N-1$. The $SU(2)$ symmetry completely fixes the $p$ dependence of the coupling constants for each $S$, that is, there should only be $N$ independent coupling constants.

An explicit decomposition of the interaction in this form, where $p$ labels a component of the angular momentum transfer, is given in Appendix \ref{app:bilinears}, but it is slightly inconvenient for our purposes. The most convenient way to implement the symmetry is to use a special basis $\lbrace M^{S,\alpha} \rbrace$ (we suppress the label $N$) for the set of Hermitian $N \times N$ matrices which has the following properties:
\begin{enumerate}
\item $S$ takes integer values from $0$ to $N-1$ and $\alpha$ takes values from $-S$ to $S$.
\item For fixed $S$, under the action $M^{S,\alpha} \rightarrow U^{\dagger}M^{S,\alpha}U$ for $U$ valued in the spin-$S_0$ representation of $SU(2)$, the $M^{S,\alpha}$ transform as a spin-$S$ representation of $SU(2)$.
\item The matrices are orthogonal under the trace norm, that is $\tr\left(M^{S,\alpha}M^{S',\beta}\right) = k\delta_{S,S'}\delta_{\alpha \beta}$ for an $S-$independent constant $k$.
\end{enumerate}
For some intuition about the $M^{S,\alpha}$ basis, we see that property (2) implies that $M^{0,0}$ is $\sqrt{k/N}$ times the $N \times N$ identity matrix and that $M^{1,\alpha}$ can be chosen to be the usual spin-$S_0$ spin matrices with $\alpha = x,y,z$. The decomposition in Fig. \ref{fig:feynman} is inconvenient because it violates property (3); in this decomposition, the $S=1$ basis matrices would be $S^z$ and $S^{\pm}$, which have less convenient orthogonality properties. We choose an unusual normalization convention where the commutation relations of $SU(2)$ are $[M^{1,\alpha},M^{1,\beta}] = \sqrt{2}i\varepsilon^{\alpha \beta \gamma}M^{1,\gamma}$ with $\varepsilon$ the Levi-Civita symbol; this implies that 
\begin{equation}
k=\frac{1}{6}N(N^2-1)
\label{eqn:k}
\end{equation} 
See Appendix \ref{app:bilinears} for an explicit construction of this basis; the matrices $M^{S,\alpha}$ are particular linear combinations of Clebsch-Gordan coefficients for fusing two spin-$S_0$ objects into a spin-$S$ object. This normalization convention is chosen because the currents $\psi^{\dagger}_{\chi} M^{1,\alpha} \psi_{\chi}$ ($\chi=L,R$ and we suppress pseudospin indices) form a representation of $\mathfrak{su}(2)_k$, giving $k$ a physical meaning. See Sec. \ref{subsec:nonAbelian} for a justification of this fact.

With this basis choice, the most general marginal interaction which is symmetric under nonchiral $SU(2)$ transformations is of the form
\begin{widetext}
\begin{equation}
H_{int} = \int dx \sum_{S=0}^{N-1} g_S \sum_{\alpha = -S}^{S} :\psi_{L,m}^{\dagger}M^{S,\alpha}_{mm'}\psi_{L,m'}:(x):\psi_{R,n}^{\dagger}M^{S,\alpha}_{nn'}\psi_{R,n'}:(x)
\label{eqn:interactionH}
\end{equation}
\end{widetext}
where we have suppressed the sums over the fermion pseudospin states. The $S=0$ and $S=1$ terms have simple physical interpretations stemming from the aforementioned explicit forms of $M^{0,0}$ and $M^{1,\alpha}$. The $S=0$ term is just a contact density-density interaction $n_L n_R$, where $n_{L/R}$ are the chiral fermion densities, while the $S=1$ term is a contact Heisenberg-type interaction $\bv{S}_L \cdot \bv{S}_R$ where $\bv{S}_{L/R}$ are the chiral $SU(2)$ pseudospin densities. See Appendix \ref{app:bilinears} for the explicit construction and proof of $SU(2)$ invariance. The Hamiltonian for the full system is then
\begin{equation}
H = H_0 + H_{int}
\end{equation}
with $H_0$ the non-interacting Hamiltonian defined in Eq. \eqref{eqn:H0}.

\section{Phase Diagram for Small $N$}
\label{sec:smallN}

\subsection{RG Procedure}

We assume that all of the $|g_S|$ are small and perform perturbative RG to second order (one loop). In the free theory, all fermion bilinears have scaling dimension 1, so all of the interactions are marginal at tree level. Using standard machinery, the perturbative RG equations for many marginal operators are known to be\cite{CardyBook}
\begin{equation}
\frac{dg_S}{dl} = -\pi \sum_{S',S''}\beta_{S',S''}^{S}g_{S'}g_{S''}
\end{equation}
where the cutoff in real space is $a_0e^{l}$ (here $a_0$ is the lattice-scale cutoff of the low-energy theory at which the bare couplings are defined) and $\beta_{S',S''}^S$ is the operator product expansion (OPE) coefficient given by the short-distance identification (written in complex coordinates)
\begin{equation}
\mathcal{O}_i(z,\bar{z}) \mathcal{O}_j(w,\bar{w}) \sim \sum_k \frac{\beta^k_{ij}\mathcal{O}_k(w,\bar{w})}{|z-w|^2}
\label{eqn:OPEdef}
\end{equation}
within correlation functions. Here, we are using a specific form of OPE where all the operators $\lbrace \mathcal{O}_i \rbrace$ involved are marginal, which is immediately applicable to our discussion. For our interactions, the OPE coefficients can be computed by Wick's theorem to be
\begin{equation}
\beta_{S',S''}^{S} = \sum_{\alpha, \beta}\frac{1}{k^2}\tr\left(\left[M^{S',\alpha},M^{S'',\beta}\right]M^{S,\gamma}\right)^2 
\label{eqn:MProductBeta}
\end{equation}
A tedious calculation, outlined in Appendix \ref{app:RGderivation}, using the explicit forms of the $M$ matrices and sum-of-product identities for the Clebsch-Gordan coefficients\cite{AngularMomentumBook} shows that 
\begin{widetext}
\begin{equation}
\beta_{S',S''}^S = -k(2S'+1)(2S''+1)\left(\begin{Bmatrix}
S & S' & S''\\
S_0 & S_0 & S_0
\end{Bmatrix}\right)^2\left(1-(-1)^{S+S'+S''}\right)^2
\label{eqn:explicitBetas}
\end{equation}
\end{widetext}
where the $\begin{Bmatrix}
S & S' & S''\\
S_0 & S_0 & S_0
\end{Bmatrix}$ is the Wigner $6j$-symbol. This form makes explicit a selection rule resulting from the symmetry properties of products of the $M$s: $\beta_{S',S''}^S$ is zero if $S+S'+S''$ is even. See Appendix \ref{app:selection} for an explanation of this selection rule in terms of Young tableaux.

Since the identity matrix commutes with all the other $M$s, $\beta^0_{S',S''}$ and $\beta^{S'}_{0,S''} = \beta^{S'}_{S'',0}$ are zero unless $S'=S''=0$. As such, to this order in perturbation theory, the $U(1)$ charge sector of the theory decouples from the pseudospin sector and, since Umklapp scattering is generally forbidden thanks to the incommensurate filling, the charge sector remains a gapless Luttinger liquid. The coupling constant $g_0$ simply changes the Luttinger parameter. We will therefore ignore the $U(1)$ sector and $g_0$ unless otherwise stated.

\subsection{Non-Abelian Bosonization}
\label{subsec:nonAbelian}

\subsubsection{Basics of Non-Abelian Bosonization}

We will use non-Abelian bosonization\cite{WittenNonAbelianBosonization} to find the strong-coupling fixed points and to determine the low-energy theories. A full review of non-Abelian bosonization is beyond the scope of this paper; we will simply define notation and briefly review the basics.

The main result of non-Abelian bosonization is that a theory of $N$ free fermions with the same Fermi velocity are equivalent to the Wess-Zumino-Witten (WZW) model $\mathfrak{u}(N)_1 = \mathfrak{u}(1) \otimes \mathfrak{su}(N)_1$. The chiral $SU(N)$ symmetry currents $J^a_{\chi}$, where $a$ labels a generator $t^a$ of $SU(N)$ and $\chi = L,R$ labels left- and right-movers, correspond to chiral fermion bilinears
\begin{equation}
J^a_{\chi}(x) \sim :\psi^{\dagger}_{m,\chi}t^a_{mn}\psi_{n,\chi}:(x)
\end{equation}
The colons indicate normal ordering and the $t^a$ generate the fundamental representation of $\mathfrak{su}(N)$. The conserved chiral currents of the $U(1)$ part of the theory are identified with the chiral total fermion density. A heuristic way to understand this identification from the CFT point of view follows from comparing operator product expansions (OPEs). Suppressing matrix indices, Wick's theorem implies that if $A$ and $B$ are matrices, then the corresponding fermion bilinears have the OPE (in complex coordinates)
\begin{widetext}
\begin{equation}
:\psi^{\dagger}_L A \psi_L:(z) :\psi^{\dagger}_L B \psi_L:(w) \sim \frac{:\psi^{\dagger}_L [A,B] \psi_L:(w)}{z-w} + \frac{\tr(AB)}{(z-w)^2 }  \label{eqn:bilinearOPE}
\end{equation}
\end{widetext}
with an analogous equation for the right-movers. With the normalization $f^{ab}_cf^{ab}_d=2N\delta_{cd}$, with $f^{ab}_c$ the structure constants of $\mathfrak{u}(N)$,  plugging in $A=t^a$ and $B=t^b$ yields the correct $\mathfrak{u}(N)_1$ OPEs
\begin{equation}
J^a_L(z) J^b_L(w) \sim \frac{i f^{ab}_cJ^c_L(w)}{z-w} + \frac{\delta_{ab}}{(z-w)^2}  
\end{equation}
More generally, Eq. \eqref{eqn:bilinearOPE} means that for any Lie subgroup $G \subset U(N)$ with generators $\tilde{t}^a$, the fermion bilinears  $\psi^{\dagger}_L\tilde{t}^a\psi_{L}$ will have the same OPEs as the symmetry currents of a WZW theory with Lie group $G$ and level $k$ equal to the embedding index $x_e$ of $G$ in $U(N)$. 

We will frequently make use of such embeddings. In order to explicitly distinguish between the currents in different subgroups $G$, define the dim$(G)$-component object $\bv{J}^{G}_{\chi}$ whose $a$th component is the current $J^a_{\chi}$, where $a$ labels a generator of $G$. In this notation the Sugawara Hamiltonian for a level-$k$ WZW theory with symmetry group $G$ is
\begin{equation}
H = \frac{1}{2(k+g)}\left(:\bv{J}^G_L \cdot \bv{J}^G_L:+:\bv{J}^G_R \cdot \bv{J}^G_R:\right)
\end{equation}
where $g$ is the dual Coxeter number of $G$.

\subsubsection{Coset construction}

Embeddings of the previously mentioned sort naturally lead to consideration of coset models; we briefly review the construction\cite{GKO}.

Consider a unitary WZW theory at level $k$ over a Lie group $G$ with a subgroup $H$, with corresponding Lie algebras $\mathfrak{h} \subset \mathfrak{g}$. Then the generators of $\mathfrak{h}$ can be written as linear combinations of  generators of $\mathfrak{g}$, so there exist currents $\bv{J}^{H}_{\chi}$ which are linear combinations of the currents $\bv{J}^{G}_{\chi}$ of the same chirality. These currents also satisfy a Kac-Moody algebra for $\mathfrak{h}$ at the level $k' = x_e k$ where $x_e$ is the embedding index of $H$ in $G$. We define the energy-momentum tensor for the coset theory $\mathfrak{g}_k/\mathfrak{h}_{k'}$ by
\begin{equation}
T_{coset} =  T_{\mathfrak{g}_k}-T_{\mathfrak{h}_{k'}}
\end{equation}
where $T_{\mathfrak{g}_k}$ and $T_{\mathfrak{h}_{k'}}$ are the energy-momentum tensors for the $\mathfrak{g}_k$ and $\mathfrak{h}_{k'}$ WZW theories respectively. The coset theory is another unitary CFT with central charge
\begin{equation}
c_{coset} = c_{\mathfrak{g}_k}-c_{\mathfrak{h}_{k'}}
\end{equation}
Importantly, the Hilbert space for the $\mathfrak{g}_k$ theory decomposes into a tensor product of the Hilbert space of the $\mathfrak{h}_{k'}$ theory and the coset theory, that is, any operator $\mathcal{O}$ in the $\mathfrak{g}_k$ theory can be written as a linear combination 
\begin{equation}
\mathcal{O} = \sum_{ij}\mathcal{O}^{\mathfrak{h}}_i \otimes \mathcal{O}^{(coset)}_j
\label{eqn:cosetOperatorDecomp}
\end{equation}
where $\mathcal{O}^{\mathfrak{h}}_i$ and $\mathcal{O}^{(coset)}_j$ are operators in the $\mathfrak{h}_{k'}$ and coset theories respectively. If $\mathcal{O}$ is a scaling operator, then its scaling dimension (conformal spin) is the sum of the dimensions (spins) of $\mathcal{O}^{\mathfrak{h}}_i$ and $\mathcal{O}^{(coset)}_j$.

A special case will be helpful later. Suppose that $\mathfrak{g}_k = \mathfrak{su}(N)_1$, $\mathfrak{h}_{k'} = \mathfrak{su}(2)_k$ with $k$ defined in Eq. \eqref{eqn:k}, and $\mathcal{O}_{L,m}$ is a chiral spin-$S$ fermion bilinear ($m=-S,...,S$ labels a component), which has scaling dimension 1. Then if we decompose $\mathcal{O}_{L,m}$ as in Eq. \eqref{eqn:cosetOperatorDecomp}, its $\mathcal{O}^{\mathfrak{su}(2)}_i$ part must have a scaling dimension less than $1$ and furthermore has to transform as a spin-$S$ field under the $\mathfrak{su}(2)_k$ algebra. This means that the $\mathcal{O}^{\mathfrak{su}(2)}_i$ part of $\mathcal{O}_{L,m}$ can only be the left-moving spin-$S$ primary $\phi^S_{L,m}$ in $\mathfrak{su}(2)_k$, i.e. 
\begin{equation}
\mathcal{O}_{L,m}\mathcal{O}_{R,m} = \phi^S_{L,m}\phi^S_{R,m}\otimes \mathcal{O}^{(coset)}
\end{equation}
for some coset operator $\mathcal{O}^{(coset)}$ with scaling dimension
\begin{equation}
\Delta_{\mathcal{O}} = 2-\frac{2S(S+1)}{k+2}
\label{eqn:scalingCoset}
\end{equation}
because $S(S+1)/(k+2)$ is the scaling dimension of $\phi^S_{L,m}$.

Before determining the phase diagram, one more notational convention is needed. A symplectic group will sometimes appear as an emergent symmetry, but the term ``symplectic group" and the notation $Sp(N)$ are used in multiple incompatible ways in the literature. In this paper, the term ``symplectic group" will always refer to the group $USp(2M)$, which is the set of $2M \times 2M$ matrices which are both unitary and preserve the symplectic form. Our notation for the Lie algebra of $USp(2M)$ is $\mathfrak{sp}(2M)$. For example, in this notation $\mathfrak{sp}(4) \approx \mathfrak{so}(5)$.

Before discussing general $N$, we analyze the cases of $N=2,3,$ and $4$ in detail. Each case will add new structure and features to the problem, but $N$ is small enough to demonstrate all of our reasoning very explicitly.

\subsection{$N=2$: Luther-Emery phase diagram}

\begin{figure}
\includegraphics[width=7cm]{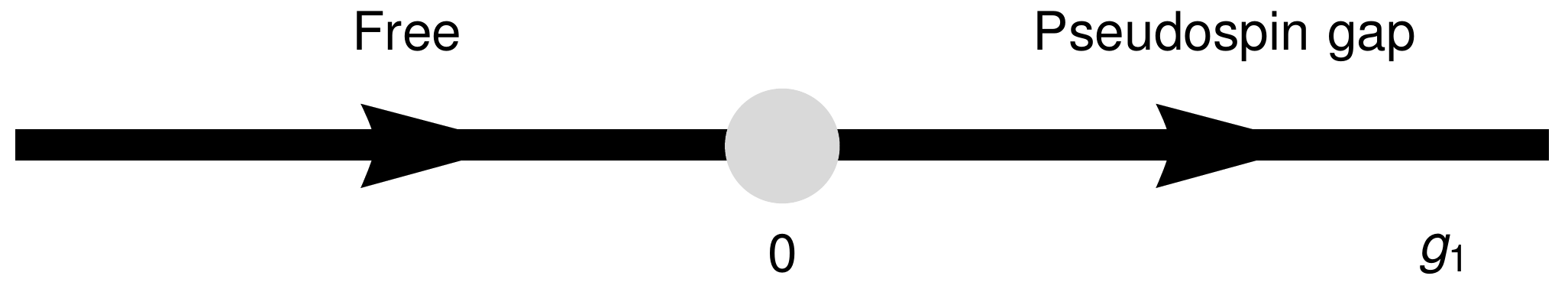}
\caption{RG flow for $N=2$. }
\label{fig:N2RG}
\end{figure}

The $N=2$ interaction Hamiltonian is simply 
\begin{equation}
H_{int} = \int dx \left(\frac{g_0}{2} n_L(x) n_R(x) + g_1 \bv{J}_L^{SU(2)}(x)\cdot \bv{J}_R^{SU(2)}(x)\right)
\end{equation} 
with $g_0$ exactly marginal and RG equation
\begin{equation}
\frac{dg_1}{dl} = 4\pi g_1^2
\end{equation}
for $g_1$. Its flow is shown in Fig. \ref{fig:N2RG}.

When $g_1 <0$ this coupling is marginally irrelevant and provides logarithmic corrections to the free-pseudospin fixed point. When $g_1 > 0$ it is marginally relevant and $\bv{J}^{SU(2)}_L \cdot \bv{J}^{SU(2)}_R$ flows to strong coupling. The latter phase is the well-known Luther-Emery phase\cite{LutherEmery} of the 1D spin-1/2 fermion chain (note that under our sign conventions, $g_0 < 0$ and $g_1>0$ when on-site interactions are attractive); strong backscattering causes the pseudospin sector to become gapped while the charge sector remains gapless. Both pseudospin-singlet CDW order at wavevector $2k_F$ and pseudospin-singlet SC have power-law correlations in this phase.

A comment on terminology: since we are studying one-dimensional physics, there is no true long-range order, only power-law correlations. We will use the terminology ``fluctuating order parameter" to describe objects which acquire such correlations since such objects can be thought of as mean-field order whose long-range order has been destroyed by quantum fluctuations.

One way to qualitatively understand this phase is as follows. Since the pseudospin sector becomes gapped, any possible fluctuating order parameters must be $SU(2)$ singlets. There are two ways to make a two-particle $SU(2)$ singlet order parameter: one in the particle-hole channel and one in the particle-particle channel. The fact that this is possible is special to $SU(2)$; particles and holes transform in conjugate representations, but representations of $SU(2)$ are self-conjugate. This means that both the singlet CDW and the singlet SC order parameters can fluctuate, and it is known that they do both fluctuate. Such arguments will be useful sanity checks in higher-$N$ cases.

\subsection{$N=3$: Two Nontrivial Phases}
\label{subsec:N3RG}

The $N=3$ RG equations are
\begin{align}
\frac{dg_1}{dl} &= 4\pi\left(g_1^2 + 5g_2^2\right)\\
\frac{dg_2}{dl} &= 24\pi g_1 g_2
\end{align}
\begin{figure}
\subfigure[ ]{
\includegraphics[width=4cm]{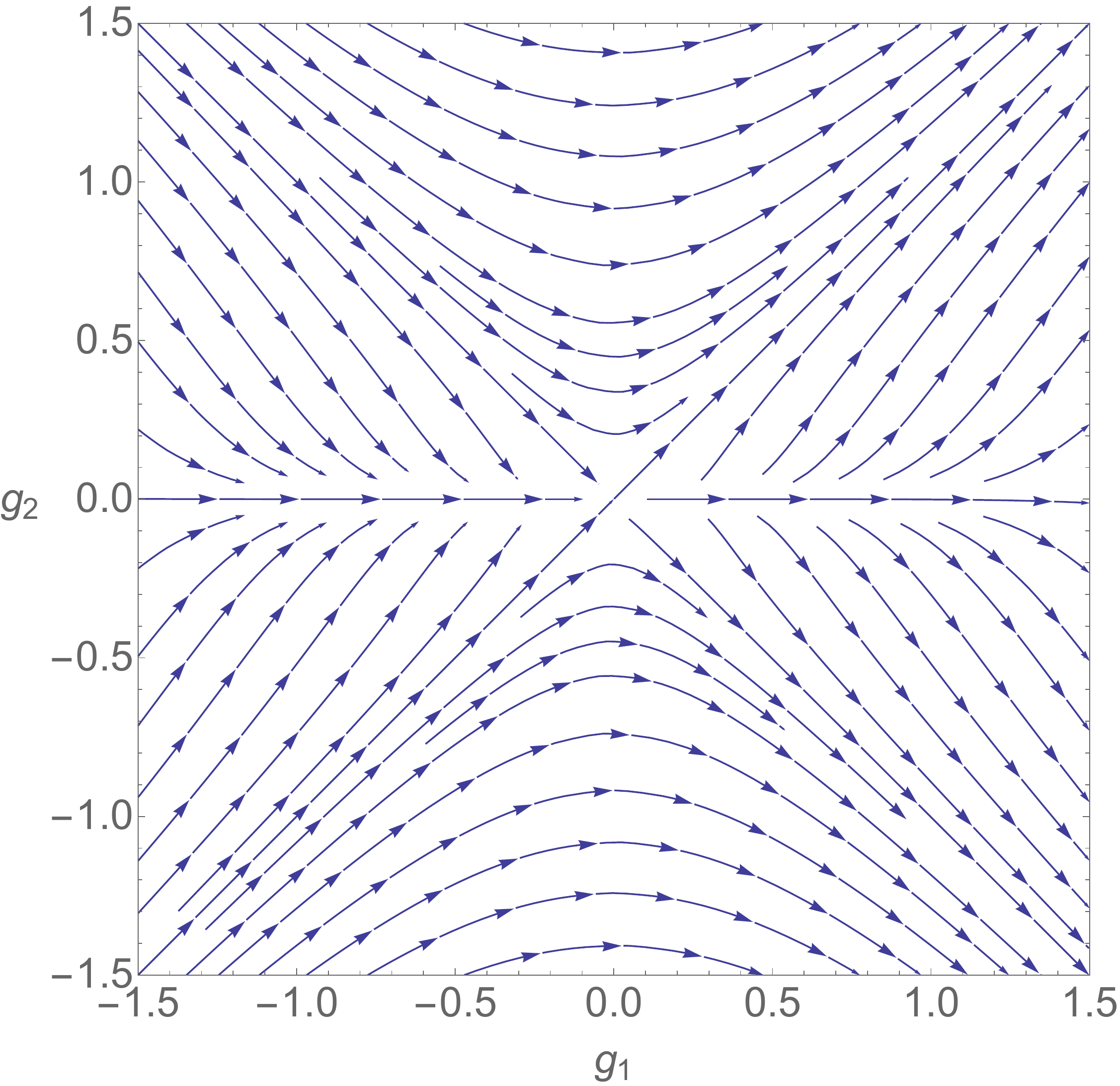}
\label{fig:N3RGFull}
}
\subfigure[ ]{
\includegraphics[width=4cm]{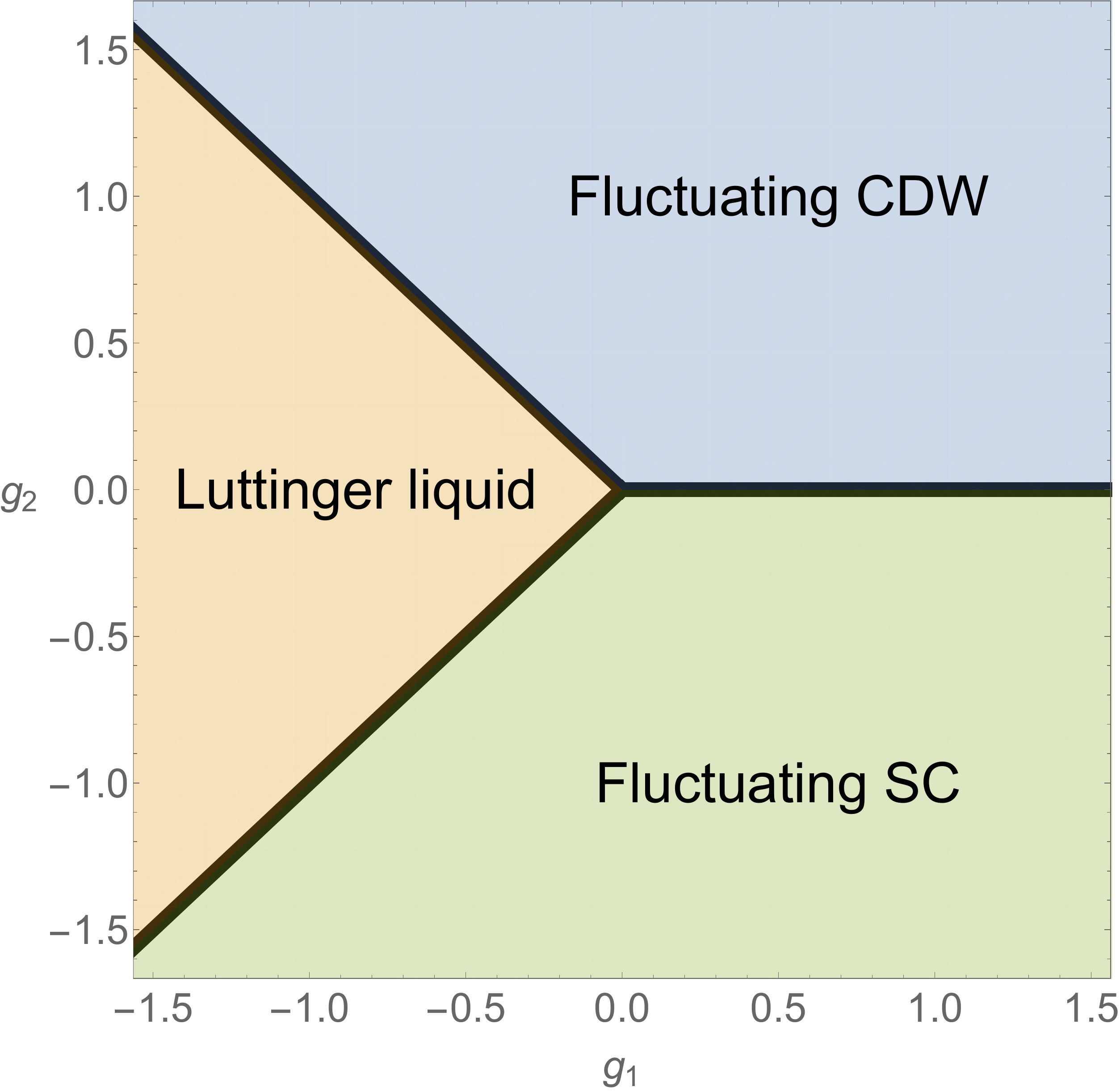}
\label{fig:N3PhaseDiagram}
}
\subfigure[ ]{
\includegraphics[width=8cm]{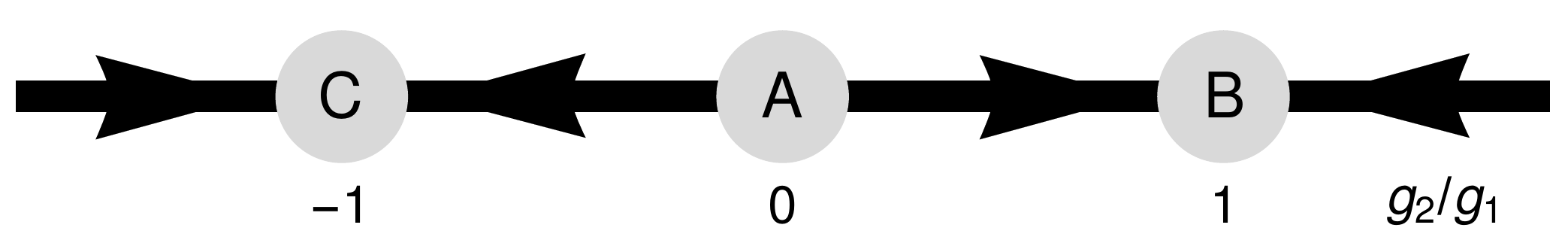}
\label{fig:N3RGRatio}
}
\caption{RG flows for $N=3$. (a) The full RG flow; the only finite-coupling fixed point is at the origin. (b) Corresponding phase diagram. (c) RG for $\tilde{g}_2 = g_2/g_1$ with $g_1>0$. Points B and C are stable fixed ``rays" corresponding to 45-degree lines in (a).}
\end{figure}
The flows in Fig. \ref{fig:N3RGFull} show that $g_1$ flows to strong coupling unless $g_1<0$ and $|g_2|<|g_1|$; if the latter occurs, both $g_1$ and $g_2$ flow to zero, and the free pseudospin fixed point is stable. In the strong-coupling case, it will be useful to define $\tilde{g}_2 = g_2/g_1$ to obtain the equation
\begin{equation}
\frac{1}{g_1} \frac{d\tilde{g}_2}{dl} = 5\tilde{g}_2\left(1-\tilde{g}_2^2\right)
\label{eqn:N3RGRatio}
\end{equation}
Clearly $\tilde{g}_2 = \pm 1$ and $g_2 = 0$ are ``fixed rays" of the RG flow, in the sense that the ratio of the coupling constants remains fixed but $g_1$ flows to strong coupling. It is easy to check by linearizing Eq. \eqref{eqn:N3RGRatio} that the fixed rays $g_2 = \pm g_1$ are stable to small changes in $\tilde{g}_2$ and the $g_2 = 0$ fixed ray is unstable; flow of this ratio is shown in Fig. \ref{fig:N3RGRatio} for $g_1>0$. The properties of the fixed points are summarized in Table \ref{table:N3phases}.

\begin{table*}
\begin{tabular}{c|c|c|c|c|c}
Label & $\tilde{g}_2$ & Stability & Symmetry of $H$ & Low-energy theory & Mean-field order \\ \hline 
A & $0$ & Unstable & $SU(2)$ & pseudospin gap & singlet CDW/SC\\
B & $1$ & Stable & $SU(3)$ & pseudospin gap & singlet CDW\\
C & $-1$ & Stable & $SO(3)$ & pseudospin gap & singlet p-wave SC
\end{tabular}
\caption{List of fixed rays and their properties for $N=3$ with $g_1>1$. The ``label" refers to Fig. \ref{fig:N3RGRatio}.}
\label{table:N3phases}
\end{table*}

What is the nature of the strong-coupling phases? By non-Abelian bosonization, the free theory is the $\mathfrak{u}(3)_1 = \mathfrak{u}(1) \otimes \mathfrak{su}(3)_1$ WZW theory. Since the $\mathfrak{u}(1)$ charge sector has decoupled, the pseudospin sector of the free theory is just $\mathfrak{su}(3)_1$. When $g_1 = g_2$, the interaction is actually fully $SU(3)$-symmetric; in the language of non-Abelian bosonization, the interaction is backscattering of the form $g \bv{J}^{SU(3)}_L \cdot \bv{J}^{SU(3)}_R$. That is, there is an emergent $SU(3)$ symmetry. When $g$ flows to strong coupling, we expect the $\mathfrak{su}(3)$ sector to be gapped; the pseudospin sector drops out of the low-energy theory entirely. 

Physically, since there is a pseudospin gap, we expect any fluctuating order parameter to be a singlet under the emergent $SU(3)$ symmetry. Since $\psi_m$ transforms under the fundamental representation of $SU(3)$, which is not self-conjugate, no particle-particle order parameter can be such a singlet. However, there is a particle-hole singlet $\psi^{\dagger}_{L,m}\psi_{R,m}$, which is, physically, the CDW order parameter. We therefore expect this phase to have fluctuating pseudospin-singlet CDW order.

Let us next consider the $g_2 = -g_1$ fixed ray, which for future purposes we will refer to as the $SO(3)$-invariant fixed ray. (The spin-1 representation of $SU(2)$ is, of course, also a representation of $SO(3)$, hence the name. Although $SO(3)$ is not an emergent symmetry, we will see that at larger odd $N$ there will be an emergent $SO(N)$ symmetry, so we choose this name to agree with the generalization.) To understand this phase, define the second-quantized operator $\hat{C}$, which is unitary at the level of the low-energy theory and acts as
\begin{align}
\hat{C}\psi_{R,m}\hat{C}^{-1} = (-1)^{m-S_0}\psi^{\dagger}_{R,-m} \nonumber\\
\hat{C}\psi^{\dagger}_{R,m}\hat{C}^{-1} = (-1)^{m-S_0}\psi_{R,-m}
\label{eqn:chiralTransform}
\end{align}
where $m = -S_0, -S_0+1,...,S_0$ and acts as the identity on the left-moving sector. Using Clebsch-Gordan coefficient identities detailed in Appendix \ref{app:bilinears}, it can be checked that
\begin{equation}
\hat{C}\psi^{\dagger}_{R,m}M^{S,\alpha}_{mn}\psi_{R,n}\hat{C}^{-1} = (-1)^{S+1}\psi^{\dagger}_{R,m}M^{S,\alpha}_{mn}\psi_{R,n}
\label{eqn:chiralMTransform}
\end{equation}
That is, $\hat{C}$ transforms the Hamiltonian at the $SU(3)$-invariant fixed ray to the Hamiltonian at the $SO(3)$-invariant fixed ray. Naively, $\hat{C}$ looks unitary, which would mean that there is an energy gap and a full $SU(3)$ symmetry at the $SO(3)$-invariant fixed ray. However, $\hat{C}$ is chiral, so this $SU(3)$ symmetry may be anomalous. As the low-energy theory suffers from the chiral anomaly, we expect any chiral symmetry to be broken in the UV, but there is no reason to expect a large perturbation to the low-energy theory. Therefore, the conclusion that there is a pseudospin gap should be robust, but the $SU(3)$ symmetry need not be.

To see what symmetry could remain in the UV, note that the $M^{S,\alpha}$ are $N \times N$ Hermitian matrices and therefore generate the chiral action of the $SU(3)$ symmetry. At the $SU(3)$ fixed ray, the nonchiral symmetry is generated by acting with the same $M^{S,\alpha}$ on both the left- and right-moving fermions. Therefore, the action of any nonchiral symmetry at the $SU(3)$ fixed ray becomes chiral at the $SO(3)$ fixed point if and only if it is generated by an $M^{S,\alpha}$ which transforms nontrivially under $\hat{C}$. Eq. \eqref{eqn:chiralMTransform} thus shows that the transformations generated by the odd-$S$ generators remain exact symmetries but those generated by the even-S generators are broken by the quantum anomaly. For $N=3$ this leaves only the $S=1$ generators, which generate $SO(3)$; therefore, the true symmetry at the fixed point should be $SO(3)$.

To get a physical understanding of the $SO(3)$ fixed point, note that $\hat{C}$ transforms density-wave order parameters into superconducting ones and vice-versa. In particular, it is easy to check that it turns the $SU(2)$-singlet CDW order parameter into the $SU(2)$-singlet SC order parameter and vice-versa. Since the CDW order parameter fluctuates in the $SU(3)$-invariant phase, the SC order parameter must fluctuate in this $SO(3)$-invariant phase while the CDW order parameter should have exponentially decaying correlations.

Our analysis so far has yielded the phase diagram of Fig. \ref{fig:N3PhaseDiagram}.

We next turn to the unstable $g_2 = 0$ fixed ray, which represents a phase transition between the CDW and the SC phases. We analyze this in a way which is slightly laborious for this particular case but will be extremely useful in more general cases.

We know that the generators of the $SU(2)$ symmetry form a representation of $\mathfrak{su}(2)_4$. Moreover, the interaction $g_1$ is exactly a product of those generators. As such, it is useful to decompose $\mathfrak{su}(3)_1 = \mathfrak{su}(2)_4 \otimes (\mathfrak{su}(3)_1/\mathfrak{su}(2)_4)$ where $\mathfrak{su}(3)_1/\mathfrak{su}(2)_4$ is a coset theory. It so happens that there is a conformal embedding of $\mathfrak{su}(2)_4$ into $\mathfrak{su}(3)_1$\cite{CFTBook}; this means that this coset theory has zero central charge and is thus trivial. But we have added a term $g_1 \bv{J}_L^{SU(2)} \cdot \bv{J}_R^{SU(2)}$ which is flowing to strong coupling; we thus expect the $\mathfrak{su}(2)_4$ theory to be gapped out. Thus we expect the strongly coupled fixed point to also have a pseudospin gap.

The fact that the phase transition appears to be gapped leaves two possibilities: either there is a first-order transition, or there is some reason that the $\mathfrak{su}(2)_4$ theory is not gapped out. In Section \ref{sec:phaseID}, we will see that our simple arguments identifying the physical character of these phases can be put on more solid ground using Abelian bosonization, and we will use those techniques to argue why one should expect a first-order transition. We defer further discussion of this phase transition to that section.

Before moving to $N=4$, a comment on the interpretation of the superconducting order parameter is in order. For $N=3$ (pseudospin-1), the two-particle singlet has a symmetric pseudospin wavefunction. Therefore, no pseudospin-singlet s-wave superconducting order parameter  can exist by Pauli statistics. However, a p-wave order parameter can exist and fluctuate.

\subsection{$N=4$: Three Nontrivial Phases}
So far we have seen quasi-one-dimensional physics appear, although the main difference between $N=2$ and $N=3$ was whether or not the singlet CDW and SC order parameters fluctuated simultaneously. However, new structure will clearly appear at $N=4$, where the non-interacting pseudospin sector is $\mathfrak{su}(4)_1$, and the level of the $\mathfrak{su}(2)$ subalgebra is $k=10$.

\begin{figure}
\centering
\subfigure[ ]{
\includegraphics[width=6cm]{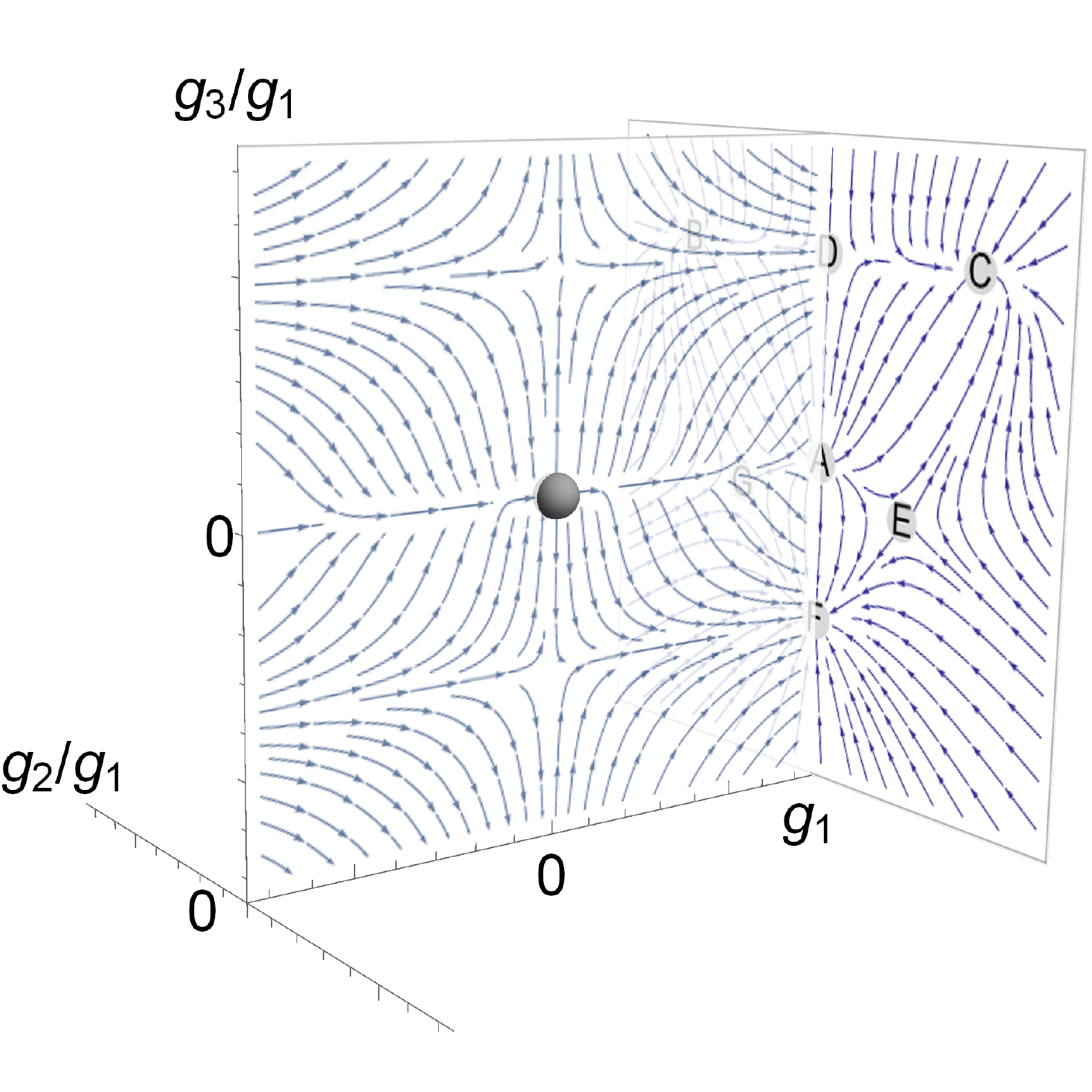}
\label{fig:N43DRG}
}
\subfigure[ ]{
\includegraphics[width=6cm]{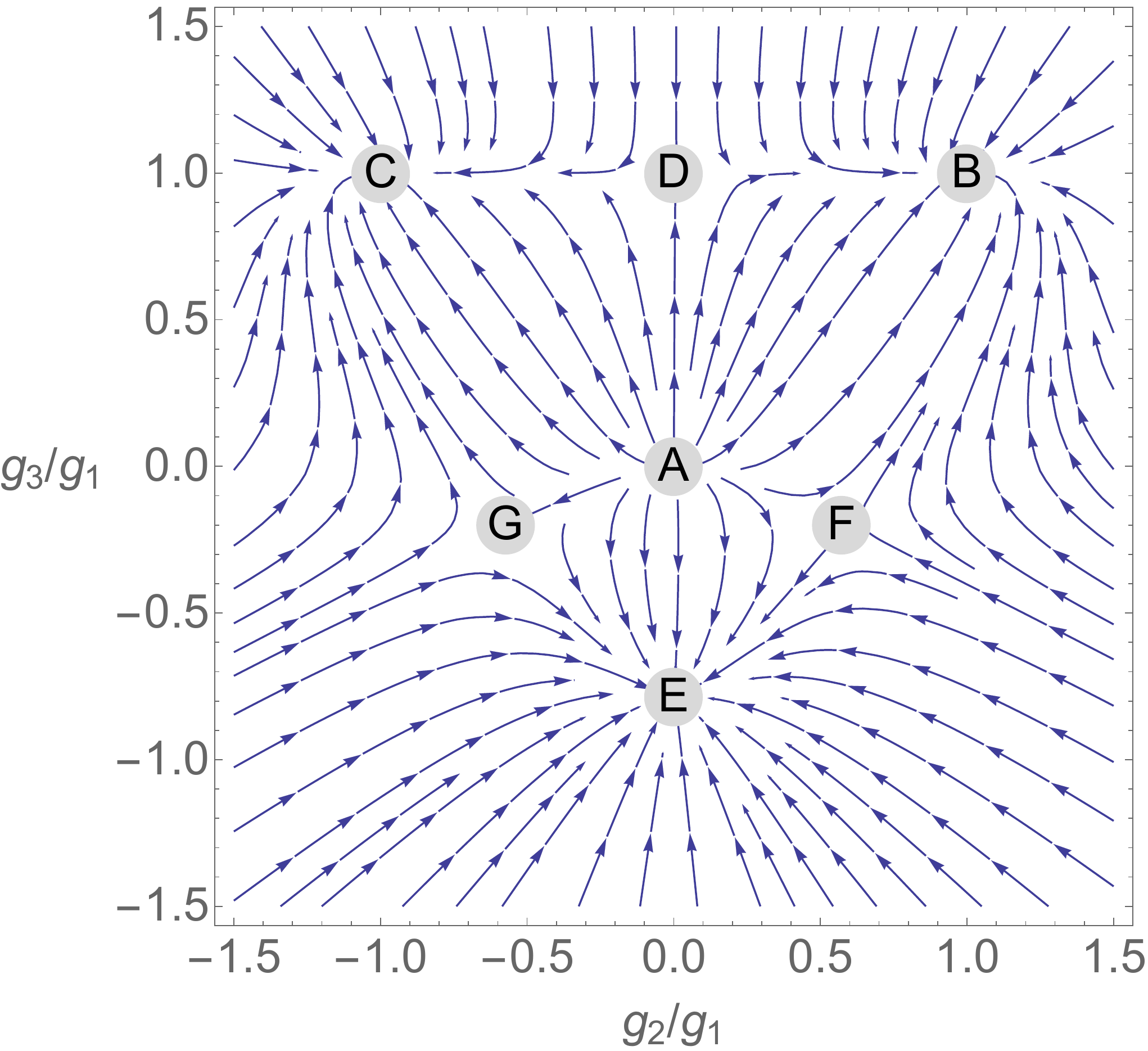}
\label{fig:N4RG}
}
\subfigure[ ]{
\includegraphics[width=6cm]{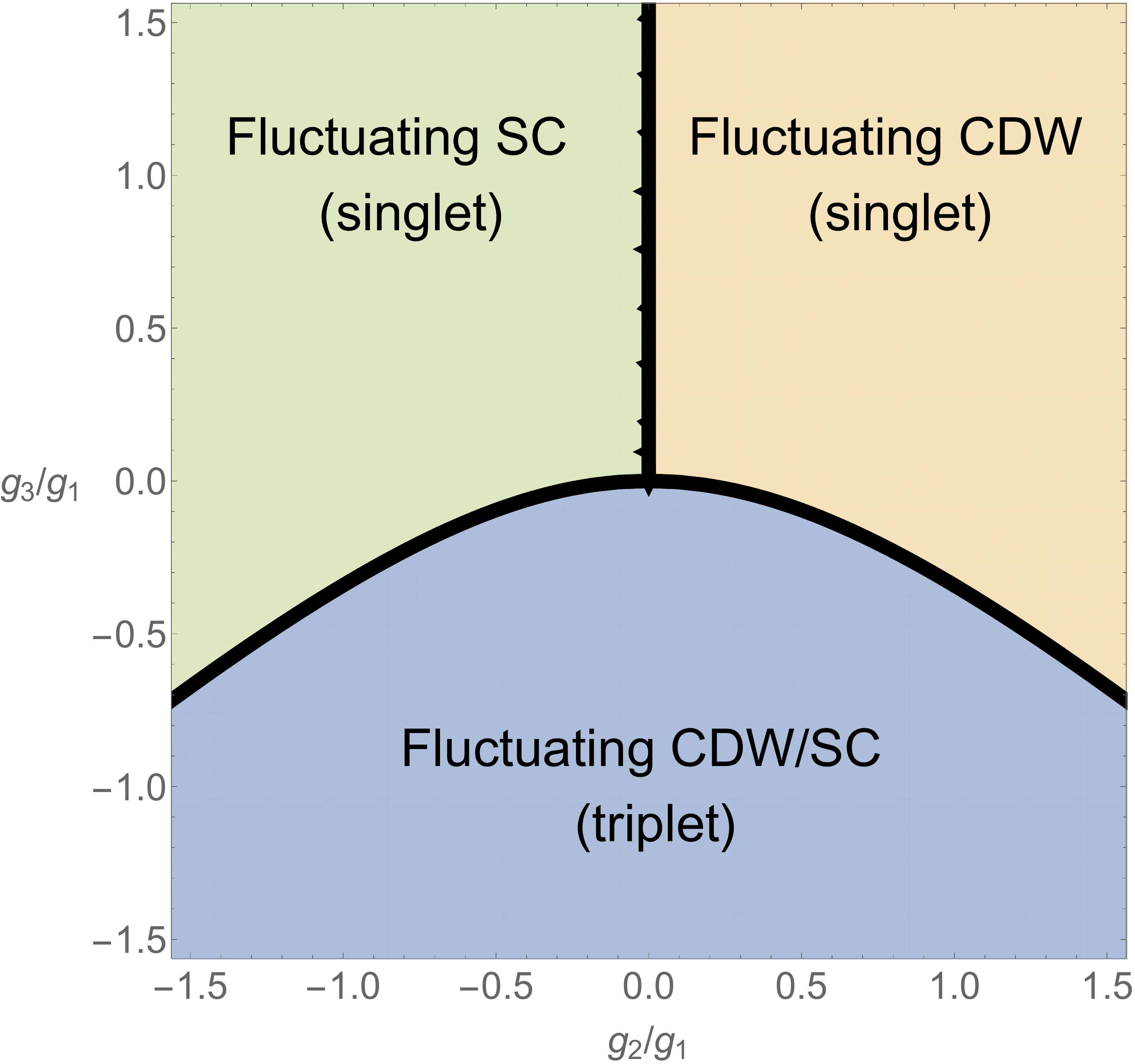}
\label{fig:N4PhaseDiagram}
}
\caption{(a) Two cuts, at $g_2 = 0$ and a schematic cut as $g_1 \rightarrow +\infty$, of the RG flow as a function of $g_1$, $g_2/g_1$, and $g_3/g_1$ for $N=4$. The sphere at the origin is the free pseudospin fixed point. (b) RG flows for the ratios of coupling constants for $N=4$ and $g_1>0$. The labeled grey circles are fixed rays where the ratio of the couplings remains fixed but all couplings become strong. (b) Phase diagram corresponding to the flows in (a).}
\end{figure}

The RG equations are
\begin{align}
\frac{dg_1}{dl} &= 4\pi\left(g_1^2+5g_2^2+14g_3^2\right)\\
\frac{dg_2}{dl} &= 4\pi\left(6g_1g_2+14g_2g_3\right)\\
\frac{dg_3}{dl} &= 4\pi\left(12g_1g_3+5g_2^2+3g_3^2\right)
\end{align}
Cuts of the flow diagram as a function of $g_1$ and the $g_{2,3}/g_1$ are shown in Fig. \ref{fig:N43DRG}, in analogy to Fig. \ref{fig:N3RGFull} for $N=3$. Focusing first on $g_1 < 0$, we see that there is a region with $g_3$ small where the free pseudospin fixed point is stable. (It is easy to check numerically that this region is stable to adding a small nonzero $g_2$). Otherwise, $g_1$ passes through zero. Although this causes $g_3/g_1$ to blow up in finite RG time, $g_3$ can still remain small and our perturbative expansion remains valid as $g_1$ changes sign; we are then reduced to studying the $g_1>0$ case. 

When $g_1>0$, it is again useful to re-analyze the equations in terms of $\tilde{g}_S = g_S/g_1$:
\begin{align}
\frac{1}{g_1}\frac{d\tilde{g}_2}{dl} &= 5\tilde{g}_2 + 14 \tilde{g}_2\tilde{g}_3 - \tilde{g}_2\left(5\tilde{g}_2^2+14\tilde{g}_3^2\right)\\
\frac{1}{g_1}\frac{d\tilde{g}_3}{dl} &= 11\tilde{g}_3 +5 \tilde{g}_2^2 + 3 \tilde{g}_3^2 - \tilde{g}_3\left(5\tilde{g}_2^2+14\tilde{g}_3^2\right)
\end{align}
The flow diagram for the $\tilde{g}_S$ with $g_1>0$ is shown both in Fig. \ref{fig:N4RG} and in Fig. \ref{fig:N43DRG} schematically located at $g_1 \rightarrow +\infty$ plane. The ``fixed points" in this diagram are, just like in Fig. \ref{fig:N3RGRatio}, actually ``fixed rays" on which the couplings grow large but have a fixed ratio.

We see clearly from the flows that there are three stable fixed rays and four unstable ones, resulting in the phase diagram in Fig. \ref{fig:N4PhaseDiagram}. It is possible to find the fixed ray couplings explicitly. The fixed rays and their properties are summarized for $g_1>0$ in Table \ref{table:N4phases}.

\begin{table*}
\begin{tabular}{c|c|c|c|c|c}
Label & $(\tilde{g}_2, \tilde{g}_3)$ & Stability & Symmetry of $H$ & Low-energy theory & Mean-field order \\ \hline 
A & $(0,0)$ & Unstable & $SU(2)$ & Ising & singlet CDW/SC\\
B & $(1,1)$ & Stable & $SU(4)$ & pseudospin gap & singlet CDW\\
C & $(-1,1)$ & Stable & $USp(4)$ & pseudospin gap & singlet SC\\
D & $(0,1)$ & Saddle & $USp(4)$ & Ising & singlet CDW/SC\\
E & $(0,-11/14)$ & Stable & $SU(2)$ & Unknown, gapless & triplet SDW/SC\\
F & $(\sqrt{41/125}, -1/5)$ & Saddle & $SU(2)$ & Unknown, gapless & singlet CDW, triplet SC/CDW\\
G & $(-\sqrt{41/125}, -1/5)$ & Saddle & $SU(2)$ & Unknown, gapless & singlet SC, triplet SC/CDW
\end{tabular}
\caption{List of fixed rays and their properties for $N=4$ with $g_1>0$. ``Label" refers to a point in Fig. \ref{fig:N4RG}, and we have suppressed the massless $U(1)$ charge sector of the low-energy theory.}
\label{table:N4phases}
\end{table*}

\subsubsection{$SU(4)$-invariant phase}

The simplest stable fixed ray is at $\tilde{g}_2 = 1$ and $\tilde{g}_3 = 1$ (point B in Fig. \ref{fig:N4RG}). As in the $N=3$ case, such a fixed ray with $g_1=g_2=g_3$ has an emergent nonchiral version of the $SU(4)$ symmetry of the non-interacting problem. As such, under bosonization, the interaction Hamiltonian is of the form $g \bv{J}_L^{SU(4)} \cdot \bv{J}_R^{SU(4)}$. Hence the pseudospin sector will gap out completely upon flowing to strong coupling. As in the $N=3$ case, any fluctuating order parameter should be an $SU(4)$ singlet, which means that it should be pseudospin-singlet CDW order.

\subsubsection{$USp(4)$-invariant phase}

The (stable) $g_S/g_1 = (-1)^{S+1}$ fixed ray (point C in Fig. \ref{fig:N4RG}) also has emergent symmetry beyond $SU(2)$. In Appendix \ref{app:bilinears}, we prove that the ten matrices $M^{1,a}$ and $M^{3,a}$, taken together, generate $USp(4) \approx SO(5)$ (it will turn out that the language $USp(4)$ is the correct generalization), and that $M^{2,a}$ transform as a 5-dimensional representation of $USp(4)$, which is the fundamental representation of $SO(5)$. Therefore the Hamiltonian is $USp(4)$-symmetric on this fixed ray, but the coupling is not simple in this language.

We can, however, understand this $USp(4)$-symmetric phase via the same chiral particle-hole transformation that we used for $N=3$. In fact, the transformation $\hat{C}$ defined in Eq. (\ref{eqn:chiralTransform}) behaves exactly the same in the $N=4$ case (with $S_0 = 3/2$) as it does for $N=3$ ($S_0 = 1$); it switches the signs of even-pseudospin couplings, thus transforming the Hamiltonian at the $SU(4)$-invariant fixed ray to that of the $USp(4)$-invariant fixed ray. Again, Eq. (\ref{eqn:chiralMTransform}) tells us that the even-$S$ generators of $SU(4)$ become anomalous, so the $SU(4)$ symmetry is broken to $USp(4)$ in the UV. We therefore expect that, like the $SU(4)$-invariant phase, the $USp(4)$-invariant fixed point is fully gapped, but has power-law singlet SC correlations rather than power-law CDW correlations.

\subsubsection{CDW/SC phase transition}

The above two phases appeared at $N=3$, but the transition between them seemed to be first-order. However, we will now show that a second-order transition is allowed (though, of course, not required) for $N=4$ by analyzing the nontrivial saddle point fixed ray with $g_2=0$ and $\tilde{g}_3=1$ (point D in Fig. \ref{fig:N4RG}). As mentioned previously, $M^{1,a}$ and $M^{3,a}$, taken together, generate $USp(4)$; in Appendix \ref{app:bilinears}, we show that the fermion bilinears that they define generate a representation of $\mathfrak{sp}(4)_1$. As such, the fixed ray coupling is actually of the form $g \bv{J}_L^{USp(4)} \cdot \bv{J}_R^{USp(4)}$. 

In the coset construction, the free theory decomposes as $\mathfrak{su}(4)_1 = \mathfrak{sp}(4)_1 \otimes (\mathfrak{su}(4)_1/\mathfrak{sp}(4)_1)$, and the fixed point interaction should cause the $\mathfrak{sp}(4)_1$ sector to gap out. This time, however, the remaining coset theory $\mathfrak{su}(4)_1/\mathfrak{sp}(4)_1$ has central charge $1/2$, that is, it is the Ising CFT. Hence the strong coupling fixed point describes a second-order, Ising-type phase transition between two pseudospin-gapped phases, one with power-law correlations of pseudospin-singlet CDW order and the other with power-law correlations of pseudospin-singlet s-wave SC order.

\subsubsection{$SU(2)$-invariant phase}

The $g_2=0$, $\tilde{g}_3 = -11/14$ fixed ray (point E in Fig. \ref{fig:N4RG}) is much more difficult to analyze because the fixed ray Hamiltonian has no additional symmetry. We can make some progress as follows.

The free spin-1 fermion currents form a representation of $\mathfrak{su}(2)_{10}$. The pseudospin sector of the free theory can be decomposed as $\mathfrak{su}(4)_1 = \mathfrak{su}(2)_{10} \otimes (\mathfrak{su}(4)_1/\mathfrak{su}(2)_{10})$ and the spin-1 currents have strictly zero correlation functions with any operator in the coset theory. In fact $\mathfrak{su}(2)_{10}$ has central charge $c=5/2$, so the coset $\mathfrak{su}(4)_1/\mathfrak{su}(2)_{10}$ has central charge $c=1/2$ and is thus the Ising CFT. If $g_3$ were zero, as at point A in Fig. \ref{fig:N4RG}, the interaction would be of the form $g \bv{J}_L^{SU(2)}\cdot \bv{J}_R^{SU(2)}$ and would flow to strong coupling. We would expect that the $\mathfrak{su}(2)_{10}$ sector would fully gap out and we would be left with a gapless Ising theory. 

However, $g_3$ is not zero at the fixed point. The corresponding operator can be decomposed into a product of the pseudospin-3 primaries in the $\mathfrak{su}(2)_{10}$ theory and an Ising primary, as in the discussion following Eq. \eqref{eqn:cosetOperatorDecomp}. However, the chiral pseudospin-3 primary $\phi^3_{L,m}$ happens to have scaling dimension $h=1$ in $\mathfrak{su}(2)_{10}$. Therefore, by Eq. \eqref{eqn:scalingCoset}, the Ising primary has dimension 0 and is trivial, so the fixed ray Hamiltonian is
\begin{equation}
H_{int} = g\left( \bv{J}_L^{SU(2)}\cdot \bv{J}_R^{SU(2)} - \frac{11}{14}\phi^3_{L,m}\phi^3_{R,m}\right)
\end{equation}

In particular, the Hamiltonian does not couple to the Ising coset theory. We therefore conclude that the low-energy theory of this phase contains the Ising CFT and is thus gapless. However, we cannot draw conclusions about the fate of the $\mathfrak{su}(2)_{10}$ sector using any tools familiar to us. Since there is RG flow, its central charge should decrease, but it is unclear if it should gap out or, for example, flow to $\mathfrak{su}(2)_{k'}$ for some $k'<k$.

\section{Identifying the Phases}
\label{sec:phaseID}

Our RG and non-Abelian bosonization pictures were very useful for understanding what fixed points are available, the spectrum, and symmetry. However, they only provided heuristic descriptions of, for example, correlation functions within each phase. To improve on that, we first build intuition using mean field theory, which is inaccurate in 1D but will prove helpful. We will then use Abelian bosonization on the fixed rays in order to extract accurate physical interpretations and calculate some correlation functions. In this section, we first explain our general techniques and conventions, then explicitly apply them to the cases $N=2$, $3$, and $4$. 

\subsection{Mean-Field Theory}

In this subsection we outline our mean-field procedure; see Appendix \ref{app:meanfield} for the details and a more careful explanation of our heuristic use of mean-field theory. 

To do mean-field theory, we can convert the coupling constant $g_1$ in the direct channel to coupling constants $g^{E}_S$ in the exchange and $g^{C}_S$ in the Cooper channels, defined as
\begin{align}
H_{int} &= \sum_S g^E_S \sum_{\alpha} \psi^{\dagger}_{L,m}M^{S,\alpha}_{mm'}\psi_{R,m'}\psi^{\dagger}_{R,n}M^{S,\alpha}_{nn'}\psi_{L,n'} \label{eqn:exchangeCoupling}\\
&= \sum_S g^C_S \sum_{\alpha} \psi^{\dagger}_{L,m}M^{(p),S,\alpha}_{mm'}\psi^{\dagger}_{R,m'}\psi_{R,n}M^{(h),S,\alpha}_{nn'}\psi_{L,n'}
\label{eqn:CooperCoupling}
\end{align} 
Here $M^{(p)}$ and $M^{(h)}$ are defined using the same conditions as the $M$ matrices but with the appropriate transformation rules under $SU(2)$ for particle-particle and hole-hole bilinears respectively.  We will show shortly that the transformations are always linear; that is, there exist $N \times N$ matrices $K^{E}$ and $K^C$ for each $N$ such that
\begin{align}
g^{E}_S &= K^{E}_{SS'}g_{S'}
\end{align}
with a similar equation for $g^C$.

Next, we perform a Hubbard-Stratonovich transformation in either the exchange or Cooper channels, integrate out the fermions, and expanding in the set of mean-field order parameters $\psi^{\dagger}_L M^{S,\alpha}\psi_R$ or $\psi^{\dagger}_LM^{(p),S,\alpha}\psi^{\dagger}_R$. At second order, all of the order parameter fields are decoupled thanks to our orthogonalization convention $\tr(M^{S,\alpha}M^{S',\beta}) = k\delta_{SS'}\delta_{\alpha \beta}$. The expansion shows that if one of the $g$ is negative, then there is a divergent susceptibility to the corresponding order, with larger $|g|$ implying a stronger instability. The details of this calculation can be found in Appendix \ref{app:meanfield}.

We now provide an explicit formula for the matrices $K$ defined in Eqs. \eqref{eqn:exchangeCoupling} and \eqref{eqn:CooperCoupling}. This is done by matching the fermion operators appearing in those equations term by term, that is,
\begin{equation}
\sum_{S',\beta} g_{S'} M^{S',\beta}_{mm'}M^{S',\beta}_{nn'} = -\sum_{S',\beta} g_{S'}^E M^{S',\beta}_{mn'} M^{S',\beta}_{nm'}
\end{equation}
Multiplying both sides by $M^{S,\alpha}_{n'm}M^{S,\alpha}_{m'n}$ for fixed $S,\alpha$ and summing on $m,m',n,n'$, the orthogonality of the $M^{S,\alpha}$ results in
\begin{equation}
g_S^E = K^{E}_{SS'}g_{S'} = -\frac{1}{k^2}\sum_{S',\beta} g_{S'} \tr\left(M^{S,\alpha}M^{S',\beta}M^{S,\alpha}M^{S',\beta}\right)
\label{eqn:Kex}
\end{equation}
By $SU(2)$ invariance this result is independent of $\alpha$.
A nearly identical computation shows that
\begin{equation}
K^{C}_{SS'} = \frac{1}{k^2} \sum_{S',\alpha} \tr\left(M^{(p),S,\alpha}M^{S',\beta}M^{(h),S,\alpha}(M^{S,\alpha})^T\right)
\end{equation}

It is also easy to show that the operator $\hat{C}$ defined in Eq. \eqref{eqn:chiralTransform} transforms
\begin{widetext}
\begin{equation}
\hat{C} \sum_{S,\alpha} g_S^E \psi^{\dagger}_{L,m}M^{S,\alpha}_{mm'}\psi_{R,m'}\psi^{\dagger}_{R,n}M^{S,\alpha}_{nn'}\psi_{L,n'} \hat{C}^{-1}
= \sum_{S,\alpha} g_S^E \psi^{\dagger}_{L,m}M^{(p),S,\alpha}_{mm'}\psi^{\dagger}_{R,m'}\psi_{R,n}M^{(h),S,\alpha}_{nn'}\psi_{L,n'}
\end{equation}
\end{widetext}
that is, it converts an operator in the exchange channel to one in the superconducting channel. But the transformation also changes the direct channel coupling constants $g_S \rightarrow (-1)^{S+1}g_S$. We conclude, then, that
\begin{equation}
K^{E}_{SS'} = K^{C}_{SS'}(-1)^{S'+1}
\end{equation}
and will therefore only explicitly list $K^{E}$.

\subsection{Abelian Bosonization}

We introduce one free chiral boson field $\phi_{m,\chi}$ ($\chi=L,R$) for each component $\psi^{\dagger}_{m,\chi}$ of chiral fermion. Our convention is
\begin{equation}
\langle \phi_{m,\chi}(x) \phi_{n,\chi}(0) \rangle = -\delta_{m,n}\log |x|
\end{equation}
We define
\begin{align}
\phi_m &= \phi_{m,L} + \phi_{m,R}\\
\theta_m &= \phi_{m,L} - \phi_{m,R}
\end{align}
which obey the commutation relations
\begin{equation}
[\phi_m(x),\partial_y \theta_m(y)] = i\delta(x-y)
\end{equation}

The corresponding bosonization identities are
\begin{align}
\psi_{m,L}^{\dagger} &\rightarrow \eta_m e^{i \phi_{m,L}}\\
\psi_{m,R}^{\dagger} &\rightarrow \bar{\eta}_m e^{-i\phi_{m,R}}\\
\sum_{\chi}:\psi^{\dagger}_{m,\chi}\psi_{m,\chi}: &\rightarrow \partial_x \phi_m
\end{align}
where $\eta_m$ and $\bar{\eta}_n$ are mutually anticommuting Klein factors which square to 1. We have dropped normalization factors. Note that the fermion operators are left unchanged under $\phi_m \rightarrow \phi_m + 2\pi l$ for $l \in \mathbb{Z}$, so we should think of $\phi_m$ as compact bosons with $\phi_m \sim \phi_m+2\pi$.

\subsection{$N=2$}

We analyze the Luther-Emery phase at $N=2$ as a familiar example before moving to the less familiar larger-$N$ cases.

The mean-field coupling constants in the exchange channel are computed using Eq. \eqref{eqn:Kex} to be
\begin{align}
K^{E} & = -\frac{1}{2}\begin{pmatrix}
1 & 3 \\
1 & -1
\end{pmatrix}
\end{align}
That is, for $g_0 = 0$, we have $g^E_0 = g^C_0 = -3g_1/2$ and $g^E_1 = g^C_1 = g_1/2$. At mean field level, there is, as expected, an instability to a singlet CDW with order parameter $\langle \psi^{\dagger}_L(x)\psi_R(x) + \text{ h. c.}\rangle$ and to singlet SC with order parameter $\langle \psi^{\dagger}_L(x)M^{(p),0}\psi^{\dag}_R(x)\rangle = \langle \psi^{\dagger}_R(x)M^{(p),0}\psi^{\dag}_L(x)\rangle$; these two orders happen to be degenerate, which is closely related to the fact that both order parameters have power-law correlations in the Luther-Emery phase. At this level of approximation, $g_0 > 0$ will break the degeneracy in favor of CDW order and $g_0<0$ will favor superconductivity, but we know from the more accurate bosonization study that this degeneracy remains, illustrating the limitations of the mean field formalism.

In Abelian bosonization, since we expect spin-charge separation it is convenient to define charge and pseudospin bosons
\begin{align}
\phi_c &= \frac{\phi_{1/2} + \phi_{-1/2}}{\sqrt{2}}\\
\phi_s &= \frac{\phi_{1/2} - \phi_{-1/2}}{\sqrt{2}}
\end{align}
which obey the same canonical commutation relations as the $\phi_m$. The compactness of $\phi_{\pm 1/2}$ implies that $\phi_{c,s}$ are not simply compact bosons; instead, $\phi_{c,s} \sim \phi_{c,s} + \sqrt{2}\pi l_{c,s}$ where $l_c$ and $l_s$ are integers of the same parity. The $g_0$ interaction term simply renormalizes the Luttinger parameter $K$ of the charge sector. The $g_1$ interaction term bosonizes to
\begin{equation}
H_{int} = -g_1 \int dx \cos \sqrt{2} \phi_s(x) 
\end{equation}
where we have made a gauge choice to project the Klein factors to the subspace $\eta_{1/2}\eta_{-1/2}\bar{\eta}_{-1/2}\bar{\eta}_{1/2} = -1$. The pseudospin sector thus becomes the sine-Gordon model, and since $g_1$ flows to strong coupling, $\phi_s$ gets pinned to $\sqrt{2}\pi l$ with $l \in \mathbb{Z}$. All values of $l$ lead to physically equivalent configurations.

Now we just need to bosonize the possible order parameters. They are
\begin{align}
\Delta_{CDW}(x) &= \sum_m e^{2ik_F x}:\psi^{\dagger}_{m,L}\psi_{m,R}: \nonumber \\
&= \eta_{1/2}\bar{\eta}_{1/2}e^{2ik_F x} e^{i\phi_c/\sqrt{2}}\cos\left(\frac{\phi_s}{\sqrt{2}}\right)\\
\Delta_{SC}(x) &= \psi^{\dagger}_{1/2,L}\psi^{\dagger}_{-1/2,R}-\psi^{\dagger}_{-1/2,L}\psi^{\dagger}_{1/2,R} \nonumber \\
&= \eta_{1/2}\bar{\eta}_{-1/2}e^{i\theta_c/\sqrt{2}}\cos\left(\frac{\phi_s}{\sqrt{2}}\right)
\end{align}
where all fields are evaluated at $x$. The pseudospin-density wave and triplet SC order parameters involve $\theta_s$ but not $\phi_s$. Since $\phi_s$ is pinned and $\partial_x \theta_s$ is its conjugate variable, the pseudospin-density wave and triplet SC order parameters have exponentially decaying correlations. On the other hand, the CDW and singlet SC order parameters fluctuate; at long distances,
\begin{align}
\langle \Delta_{CDW}(x)\Delta_{CDW}^{\ast}(0) \rangle &\sim \frac{1}{|x|^{1/K}}\\
\langle \Delta_{SC}(x)\Delta_{SC}^{\ast}(0) \rangle &\sim \frac{1}{|x|^{K}}
\end{align}

These simultaneously fluctuating order parameters, together with the spin gap, are a hallmark of the Luther-Emery phase.

\subsection{$N=3$}

For the mean-field analysis, we find
\begin{equation}
K^{E} = -\frac{1}{3} \begin{pmatrix}
1 & 3 & 5\\
1 & 3/2 & -5/2\\
1 & -3/2 & 1/2
\end{pmatrix}
\end{equation}

We first consider $g_0=0$. At the $SU(3)$-symmetric flow, the most negative coupling constant is pseudospin-singlet CDW order. At the $SO(3)$-symmetric flow, pseudospin-singlet superconductivity $\langle \psi^{\dagger}_L(x) M^{(p),0}\psi^{\dagger}_R(x)\rangle$ has the most negative coupling constant. 
Both are degenerate at the $g_2=0$ fixed ray. We thus expect a phase transition between fluctuating CDW and fluctuating singlet SC orders.

Now let us add $g_0 \neq 0$. At mean-field level, $g_0$ changes the location of the transition. In the bosonization language, there is spin-charge separation; the naive effect of a nonzero $g_0$ is simply to change the Luttinger parameter of the charge sector. Deep in a phase this merely distinguishes the power laws of correlation functions of the two order parameters. However, this distinction suggests that $g_0$ modifies the energies of the two phases relative to one another, and since the phase transition seems to be first order this may indeed modify the location of the phase transition.

To check this in Abelian bosonization, we define one charge and two pseudospin bosons
\begin{align}
\phi_c &= \frac{\sum_m \phi_m }{\sqrt{3}}\\
\phi_{s1} &= \frac{\phi_{1} - \phi_{-1}}{\sqrt{2}}\\
\phi_{s2} &= \frac{\phi_{1} + \phi_{-1}-2\phi_0}{2}
\end{align}
These fields mutually commute. Again there is pseudospin-charge separation and the only effect of $g_0$ is to renormalize the Luttinger parameter $K$ of the charge sector. Compactness of the $\phi_m$ results in compactifications of $\phi_c$, $\phi_{s1}$ and $\phi_{s2}$ generated by the identifications $(\phi_c, \phi_{s1}, \phi_{s2}) \sim (\phi_c + 2\sqrt{3}\pi, \phi_{s1}, \phi_{s2}) \sim (\phi_c, \phi_{s1} + 2\sqrt{2}\pi, \phi_{s2}) \sim (\phi_c + 2\pi/\sqrt{3}, \phi_{s1} +\sqrt{2}\pi, \phi_{s2} + \pi) $.

Analyzing the interaction for general values of $g_1$ and $g_2$ is challenging, but it is straightforward on the stable fixed rays, which, as before, we refer to as the $SU(3)$ ($g_2=g_1$) and $SO(3)$ ($g_2 = -g_1$) fixed rays. The pseudospin Hamiltonians are
\begin{align}
H_{int,SU(3)} &= -g\int dx \left(\cos \sqrt{2}\phi_{s1} + 2\cos \phi_{s2} \cos \left(\frac{\phi_{s1}}{\sqrt{2}}\right)\right)\\
H_{int,SO(3)} &= g\int dx \left(\cos \sqrt{2}\phi_{s1} - 2\sin \theta_{s2} \sin \left(\frac{\phi_{s1}}{\sqrt{2}}\right)\right)
\end{align}
where we have chosen three independent Klein factor projections and all fields inside the integrals are evaluated at $x$. The appearance of sines instead of cosines in the $SO(3)$ Hamiltonian results from the Klein factors and the odd number of fermion flavors and, as we will see, it is very important.

These Hamiltonians are unfrustrated. In the $SU(3)$ phase, $\phi_{s1}$ and $\phi_{s2}$ are pinned to $\sqrt{2} \pi l_1$ and $\pi l_2$ respectively, where $l_1$ and $l_2$ are integers of the same parity. All such configurations are physically identical. In the $SO(3)$ phase, $\phi_{s_1}$ and $\theta_{s2}$ are pinned to $\sqrt{2}\pi (l_1+1/2)$ and $\pi (l_2+1/2)$ where $l_1$ and $l_2$ again have the same parity. 

To understand what the phases do physically, we bosonize the pseudospin-singlet order parameters:
\begin{align}
\Delta_{CDW}^{S=0} &= e^{2ik_F x} e^{i\phi_c/\sqrt{3}+i\phi_{s2}/3}\eta_1\bar{\eta}_1\left(2\cos \left(\frac{\phi_{s1}}{\sqrt{2}}\right) + e^{-i\phi_{s2}}\right)\\
\Delta_{SC}^{S=0} &= e^{i\theta_c/\sqrt{3}+i\theta_{s2}/3}\eta_1\bar{\eta}_1\left(-2i\sin \left(\frac{\phi_{s1}}{\sqrt{2}}\right) + e^{-i\theta_{s2}}\right)
\end{align}
Since $\phi_{s1}$ is always pinned and $\phi_{s2}$ ($\theta_{s2}$) is pinned in the $SU(3)$ ($SO(3)$) phase, we see that singlet CDW (SC) order has power-law decay and SC (CDW) order has exponential decay. The long-distance power laws are
\begin{align}
\langle \Delta_{CDW}^{S=0}(x)\Delta_{CDW}^{S=0 \ast}(0) \rangle &\stackrel{SU(3)}{\sim} \frac{1}{|x|^{2/(3K)}}\\
\langle \Delta_{SC}^{S=0}(x)\Delta_{SC}^{S=0\ast}(0) \rangle &\stackrel{SO(3)}{\sim} \frac{1}{|x|^{2K/3}}
\end{align}

For higher-spin channels, $SU(2)$ invariance allows us to only check the $m=0$ component of the higher-spin order parameters. The spin-density wave (SDW) order parameters bosonize as follows:
\begin{align}
\Delta_{SDW}^{S=1} &\propto e^{i\phi_c/\sqrt{3}+\phi_{s2}/3}\eta_1 \bar{\eta}_1\sin \left(\frac{\phi_{s1}}{\sqrt{2}}\right)\\
\Delta_{SDW}^{S=2} &\propto e^{i\phi_c/\sqrt{3}+\phi_{s2}/3}\eta_1 \bar{\eta}_1\left(\cos \left(\frac{\phi_{s1}}{\sqrt{2}}\right) - e^{i\phi_{s2}}\right)
\end{align}
In the $SU(3)$ phase, $\phi_{s1}$ and $\phi_{s2}$ are both pinned to zero, so both order parameters are also pinned to zero. In the $SO(3)$ phase, $\theta_{s2}$ is pinned, causing both of these order parameters to have exponentially decaying correlations. The higher-spin SC order parameters are also either pinned to zero or decay similarly. The conclusion is that, as expected, only the pseudospin-singlet CDW (SC) order parameter has power-law correlations in the $SU(3)$ ($SO(3)$) phase.

Remarkably, these results are in accordance with the intuition gained from mean field theory. The channel with the most negative coupling constant has power-law fluctuations, while all others have exponentially decaying correlations.

\subsubsection{Comparison to non-Abelian results}

Notice that $\phi_{s1}$ is pinned to physically inequivalent values in the two phases. In particular, if there is an externally-enforced boundary between these two phases, $\phi_{s1}$ must change by a half-integer multiple of its compactification length $\sqrt{2}\pi$. The interpretation can be understood as follows. Clearly $\partial_x \phi_{s1}$ is proportional to the density of $S_z$. In particular locally adding a fermion with $S_z = +1$ corresponds to adding a $2\pi$ kink in $\phi_{1}$; this means that there is a $\sqrt{2}\pi$ kink of $\phi_{s1}$. Hence a $\sqrt{2}\pi$ kink in $\phi_{s1}$ corresponds to a localized change in spin by 1 unit. We instead have a $\pi/\sqrt{2}$ kink, so there must be a half-integer spin trapped at the boundary despite the system being built out of integer pseudospins. We conclude that the two phases are topologically distinct.

However, non-Abelian bosonization (see Sec. \ref{subsec:N3RG}) indicated that at the phase transition ($g_2 =0$), the low-energy theory should have central charge 0 and thus be gapped. There are therefore two possibilities:
\begin{itemize}
\item The transition at $g_2=0$ is first order.
\item The transition at $g_2=0$ is continuous, and there is a topological obstruction to gapping out $\mathfrak{su}(2)_4$ using a $\mathbf{J}_L \cdot \mathbf{J}_R$ interaction.
\end{itemize}
We cannot rule out the second possibility except to say that we have found no evidence supporting it. 
In the absence of numerical evidence, we suggest that the transition is first order.

\subsection{$N=4$}

Starting with mean field again, we find
\begin{equation}
K^{E} = -\frac{1}{4} \begin{pmatrix}
1 & 3 & 5 & 7\\
1 & 11/5 & 1 & -21/5\\
1 & 3/5 & -3 & 7/5\\
1 & -9/5 & 1 & -1/5	
\end{pmatrix}
\end{equation}

At mean field level, the leading instabilities are as follows when $g_0=0$. At the $SU(4)$- and $USp(4)$-invariant fixed points, CDW and singlet SC orders respectively have the most negative coupling constants, so we expect physics similar to $N=3$. The fixed point without emergent symmetry ($g_2 = 0, g_3 = -11/14g_1$) has degenerate pseudospin-triplet SDW order and pseudospin-triplet p-wave superconductivity. The physical picture of this phase should then be of fluctuations of both of these order parameters. Both order parameters would spontaneously break $SU(2)$ symmetry if they developed; therefore it makes sense that the pseudospin sector could remain gapless due to fluctuating Goldstone modes.

The effect of a nonzero $g_0$ is similar to that of $N=3$; again at mean-field level it modifies the location of the phase transition. However, if the transition between the $SU(4)$- and $USp(4)$-invariant phases is second-order (which is allowed for $N$ even), we expect that $g_0$ will not significantly modify the phase transition.

For the $SU(4)$- and $USp(4)$-invariant phases, the Abelian bosonization analysis is very similar to that for $N=3$. We use the fields
\begin{align}
\phi_c  &= \frac{\sum_m \phi_m}{2}\\
\phi_{s1} &= \frac{\phi_{1/2}-\phi_{-1/2}}{\sqrt{2}}\\
\phi_{s2} &= \frac{\phi_{1/2}+\phi_{-1/2}-\phi_{3/2}-\phi_{-3/2}}{2}\\
\phi_{s3} &= \frac{\phi_{3/2}-\phi_{-3/2}}{\sqrt{2}}
\end{align}
Bosonizing the fixed point Hamiltonians produces, after setting Klein factor conventions,
\begin{widetext}
\begin{align}
H_{int,SU(4)} &= -g\int dx \left(\cos(\sqrt{2}\phi_{s1})+\cos(\sqrt{2}\phi_{s3}) + 4\cos\left(\frac{\phi_{s1}}{\sqrt{2}}\right)\cos\left(\frac{\phi_{s3}}{\sqrt{2}}\right)\cos(\phi_{s2})\right)\\
H_{int,USp(4)} &= -g\int dx \left(\cos(\sqrt{2}\phi_{s1})+\cos(\sqrt{2}\phi_{s3}) + 4\cos\left(\frac{\phi_{s1}}{\sqrt{2}}\right)\cos\left(\frac{\phi_{s3}}{\sqrt{2}}\right)\cos(\theta_{s2})\right)
\end{align}
\end{widetext}
Again both Hamiltonians are unfrustrated, and the difference between the two phases is whether $\phi_{s2}$ or $\theta_{s2}$ is pinned. It is easy to check by bosonizing the order parameters that when $\phi_{s2}$ ($\theta_{s2})$ is pinned, the CDW (singlet SC) order parameter acquires power-law correlations
\begin{align}
\langle \Delta_{CDW}^{S=0}(x)\Delta_{CDW}^{S=0 \ast}(0) \rangle &\stackrel{SU(4)}{\sim} \frac{1}{|x|^{1/(2K)}}\\
\langle \Delta_{SC}^{S=0}(x)\Delta_{SC}^{S=0\ast}(0) \rangle &\stackrel{USp(4)}{\sim} \frac{1}{|x|^{K/2}}
\end{align}
There is a crucial qualitative difference between $N=3$ and $N=4$: for $N=4$, both $\phi_{s1}$ and $\phi_{s2}$ are pinned to the same set of (physically equivalent) values in both phases. This means that, unlike for $N=3$, there are no topologically protected, fractionalized edge states between these two phases. This is expected; since the onsite fermion number is not fixed, the fermions should be thought of as transforming in the fundamental representation of $USp(4) \subset SU(4)$, a symmetry which is preserved at both the CDW and SC fixed points. Being simply connected, $USp(4) \approx Spin(5)$ has no projective representations and thus there can be no fractionalization of the full symmetry. By contrast, for $N=3$, the fermions carry the fundamental of $SO(3)$, which can fractionalize into spinor representations.

The phase without emergent symmetry is unfortunately very difficult to analyze using Abelian bosonization. Even assuming that perturbative RG yielded the correct value for the ratios of couplings on the fixed ray, which need not be the case since the flow is to strong coupling, the cosine terms that appear do not all commute, so there is no simple ``pinning" picture at strong coupling. We therefore cannot confirm our mean field intuition about this peculiar phase and leave further investigation to future work.

\section{Phase Diagram for General $N$}
\label{sec:generalN}

Unfortunately, the fixed ray structure is hard to visualize for $N>4$ due to the large parameter space. We can make some exact statements for general $N$; together with example calculations and numerics done at small $N$, this is enough to guess the key features of the phase diagram at all $N$.

Before discussing the results, we briefly explain the nature of our numerical work. We evaluated Eq. \eqref{eqn:explicitBetas} numerically in order to obtain the RG equations, which were then rewritten as a function of the $\tilde{g}_S$ and solved numerically in order to obtain the full set of fixed rays. The stability of the fixed rays was evaluated by numerically linearizing the RG equations for $\tilde{g}_S$ about the fixed ray, writing $d\delta\tilde{g}_S/dl \approx A_{SS'}\delta \tilde{g}_{S'}$, where $\delta \tilde{g}_S$ is the difference between $\tilde{g}_S$ and its fixed ray value. The fixed ray is stable if and only if all of the eigenvalues of $A$ are negative; we diagonalized $A$ numerically. The fixed point structure was obtained numerically in this way for all $N \leq 8$. We also calculated $K^E$ from Eq. \eqref{eqn:Kex} by numerically generating the $M^{S,\alpha}$ using the relation to Clebsch-Gordan coefficients (which can be generated algorithmically by standard techniques) detailed in Appendix \ref{app:bilinears}.

As a first general statement, using Eq. (\ref{eqn:explicitBetas}), it is straightforward to show that the RG equation for $g_1$ is always of the form
\begin{equation}
\frac{dg_1}{dl} = \frac{2\pi}{3}\sum_S S(S+1)(2S+1) g_S^2
\end{equation}
We conjecture that, as in $N=3$ and $N=4$, there is a region where the pseudospin sector can still flow to the free fixed point when $g_1<0$, occurring when the $|\tilde{g}_S|$ are sufficiently small; in this regime, all the $|g_S|$ for $S>1$ decrease more rapidly than $|g_1|$ does. Otherwise, unless there is fine-tuning, the system will generically flow to large positive $g_1$, and the system should be analyzed using fixed rays in the same way as at small $N$.

\subsection{$SU(N)$-Invariant Phase}

Using the completeness of the Clebsch-Gordan coefficients, it can be shown that $\sum_{S',S''}\beta^S_{S',S''} = -2Nk$ for all $S$. Hence, there is a fixed ray with $g_S = g$ for all $S > 0$, and the flow is to strong coupling if $g > 0$. The existence is rigorous; we conjecture based on the numerical evidence discussed above that this fixed ray is stable.

On this fixed ray, the system has a nonchiral $SU(N)$ symmetry and the corresponding interaction, when bosonized, is of the form $g \bv{J}_L^{SU(N)} \cdot \bv{J}_R^{SU(N)}$. Hence we expect the interaction to gap out the $\mathfrak{su}(N)_1$ sector.

To understand the nature of this phase, we use similar arguments to before. Since the $\mathfrak{su}(N)$ sector is gapped out, we expect the fluctuating order parameter to be an $SU(N)$ singlet. This can only happen (for fermion bilinears) in the particle-hole channel because the fundamental representation of $SU(N)$ is not self-conjugate for $N>2$. We therefore expect the leading mean-field instability to be the pseudospin-singlet density wave (exchange) channel, which is confirmed by our numerical calculations of $K^E$. 
This phase should thus have power-law correlations of the CDW order parameter (where the power depends on $N$, see Sec. \ref{sec:largeN}). These correlations were checked explicitly in Abelian bosonization for $N \leq 6$ by generalizing the method in Section \ref{sec:phaseID}.

\subsection{Odd $N$}

In addition to the $SU(N)$-invariant fixed ray, there is always additional structure in the phase diagram. By the selection rule present in the OPE coefficents in Eq. (\ref{eqn:explicitBetas}), the number of $g_S$ with even $S$ has the same parity on both sides of the RG equation. Hence the RG equation is symmetric under $g_S \rightarrow (-1)^{S+1}g_S$, so the existence of the $SU(N)$-invariant fixed ray implies the existence of a fixed ray at $g_S/g_1 = (-1)^{S+1}$. Moreover, the chiral particle-hole transformation Eq. \eqref{eqn:chiralTransform} relates these two fixed rays at the level of the low-energy theory. This transformation causes the even-$S$ generators of the $SU(N)$ symmetry to become anomalous, and it interchanges particle-hole and particle-particle order parameters. This is true for all $N$.

Apart from the existence of these two fixed rays, however, the behavior of the phase diagram depends strongly on the parity of $N$, with odd $N$ being simpler. We first focus on this simpler case. In fact, $N=3$ contains almost all the physics of the general case for odd $N$. Our numerical solution of the RG equations finds that $\tilde{g}_S=1$ and $\tilde{g}_S = (-1)^{S+1}$ are the \textit{only} stable fixed rays . This latter fixed ray has $SO(N)$ symmetry; in fact, we prove in Appendix \ref{app:bilinears} that for odd $N$, the $M^{S,\alpha}$ for odd $S$ form the fundamental representation of $\mathfrak{so}(N)$, and that the corresponding chiral fermion currents form a representation of $\mathfrak{so}(N)_2$ for $N>3$. ($N=3$ is exceptional, forming $\mathfrak{so}(3)_4$ due to the isomorphism of the Lie algebras $\mathfrak{so}(3)$ and $\mathfrak{su}(2)$.) We also conjecture that the $SO(N)$-invariant fixed ray has power-law correlations of spin-singlet SC order (where the power again depends on $N$, see Sec. \ref{sec:largeN}). This was checked in Abelian bosonization for $N=5$; the treatment is completely analogous to $N=3$ and $N=4$.

Moreover, our numerical solution of the RG equations always shows that there is an unstable fixed ray with $g_S = 0$ for even $S$ and $g_S/g_1 = 1$ for odd $S$, analogous to the $g_2 = 0$ fixed point at $N=3$. Naively this might mark a continuous transition between an $SU(N)$-invariant phase and an $SO(N)$-invariant phase. But since the only couplings which appear involve currents in the fundamental representation of $\mathfrak{so}(N)$, the interaction is the marginally relevant coupling $g \bv{J}_L^{SO(N)} \cdot \bv{J}_R^{SO(N)}$. The fixed point at strong coupling should be described by $\mathfrak{su}(N)_1/\mathfrak{so}(N)_2$, which can be checked to have central charge 0; this is a known conformal embedding\cite{BaisConformalEmbeddings}. For the same reasons as at $N=3$, we conjecture that the transition between these phases is first-order.

\subsection{Even $N$: $USp(N)$-Invariant Phase and Parafermions}

When $N$ is even, we conjecture based on the numerical solution of the RG equations for $N \leq 8$ that the phase structure is similar to that of $N=4$. That is, in addition to the $SU(N)$-invariant phase, there is a $USp(N)$-invariant phase and a phase which has no symmetry beyond the $SU(2)$ symmetry we imposed. We focus on the former in this section.

As in the odd $N$ case, the RG equations are symmetric under $g_S \rightarrow (-1)^{S+1}g_S$, so there is (rigorously) always a fixed ray at $g_S/g_1 = (-1)^{S+1}$, which we conjecture to be stable. We prove in Appendix \ref{app:bilinears} that the $M^{S,\alpha}$ for odd $S$ generate the fundamental representation of $USp(N)$. By the selection rules resulting from Eq. \eqref{eqn:explicitBetas}, we also see that the OPE of an odd-pseudospin fermion current with an even-pseudospin current produces only even-pseudospin currents. Therefore, this phase is fully $USp(N)$-invariant.

To understand this phase, we can again use the operator $\hat{C}$ appearing in Eq. (\ref{eqn:chiralTransform}) with the appropriate value of $S_0$. Eq. \eqref{eqn:chiralMTransform} holds for any $N$, so as in the $N=4$ case, the $USp(N)$-invariant phase should have a full pseudospin gap and power-law singlet s-wave superconducting correlations. This was checked by numerical mean field calculations using $K^E$ for $N \leq 8$, which show that singlet s-wave superconductivity is the leading instability, and Abelian bosonization for $N\leq 6$.

We can also consider the phase transition between the $SU(N)$-invariant phase and the $USp(N)$-invariant phase; since at $g_S/g_1=1$ for odd $S$ and $g_S=0$ for even $S$ the system is invariant under $\hat{C}$, such a fixed point always exists. We know that the odd-pseudospin matrices generate $USp(N)$, and we have computed in Appendix \ref{app:bilinears} that the odd-pseudospin fermion bilinears generate a representation of $\mathfrak{sp}(N)_1$. We can then conjecture that if the transition is second order, then it is described by $\mathfrak{su}(N)_1/\mathfrak{sp}(N)_1$.

To understand this theory, we simply note that $\mathfrak{su}(N)_1 = \mathfrak{u}(N)_1/\mathfrak{u}(1)$, so the pseudospin sector is described by $(\mathfrak{u}(N)_1/\mathfrak{u}(1))/\mathfrak{sp}(N)_1$. Switching the order of the coset procedure (which is valid because the generator of the $\mathfrak{u}(1)$ subalgebra commutes with the generators of $\mathfrak{sp}(N)_1$), we obtain $(\mathfrak{u}(N)_1/\mathfrak{sp}(N)_1)/\mathfrak{u}(1)$. But $\mathfrak{u}(N)_1/\mathfrak{sp}(N)_1 = \mathfrak{su}(2)_{N/2}$ for even $N$, so our phase transition is described by the $\mathfrak{su}(2)_{N/2}/\mathfrak{u}(1)$ theory, which describes $\mathbb{Z}_{N/2}$ parafermions.

Although this second-order phase transition is consistent with our results, we cannot rule out the possibility of first-order phase transitions appearing instead. In fact, it is quite possible that, much like the quantum rotor model, for some values of $N$ this fixed point is actually the multicritical end of a line of first-order transitions.

\subsection{Even $N$: $SU(2)$-Invariant Phase}

Again based on the numerical solution to the RG equations for all $N \leq 8$, we conjecture that there is always another stable fixed ray for even $N$ at $g_S = 0$ for $S$ even and some particular but non-generic (and not all positive) values of $\tilde{g}_S$ for $S$ odd. Unfortunately, the analysis of this fixed point is even more challenging than for $N=4$ for two reasons. First, the coset theory $\mathfrak{su}(n)_1/\mathfrak{su}(2)_k$ has central charge
\begin{equation}
c_{coset} = \frac{(N-3)(N-2)(N-1)(N+2)}{N^3-N+12}
\end{equation}
which is not an easily identifiable theory for $N>4$. Second, it is merely a coincidence that for $N=4$, the field $\phi^3_L \phi^3_R$ has scaling dimension 2 in $\mathfrak{su}(2)_{10}$. This coincidence allowed us, using Eq. \eqref{eqn:scalingCoset}, to say that the coset theory did not flow as the fixed ray couplings grow large. In general, the interaction at the fixed point will not generally live only in the $\mathfrak{su}(2)_k$ theory; although some spin-$S$ term may happen to have scaling dimension 2, other operators are typically present, so the coset theory flows as well. 

However, based on our mean-field procedure and numerical calculations of $K^E$, we conjecture that this phase has, as at $N=4$, fluctuating pseudospin-triplet CDW and $p$-wave SC orders, and should, correspondingly, be gapless in the pseudospin sector. As an additional piece of evidence, if the fixed ray indeed always has $g_S = 0$ for $S$ even (as it does at our level of approximation for $N \leq 8$), then at the level of the low-energy theory the pseudospin part of the theory is invariant under $\hat{C}$. This means that the triplet CDW order parameter has power-law correlations if and only if the triplet SC order parameter does as well.

\section{$SU(2)$-Breaking Perturbations}
\label{sec:symmbreak}

Recall that the whole point of our mapping from a wire to $\mathbb{R} \times S^2$ geometry was to restore magnetic translation symmetry, which is broken in a wire, while also changing the group structure of magnetic translation symmetry to $SU(2)$. In order for our results to relate to real wires, we therefore need to add $SU(2)$-breaking perturbations. In this section, we give some qualitative arguments about what happens when $SU(2)$ symmetry is broken.

Recall that in the $\mathbb{R} \times D^2$ geometry in symmetric gauge, single-particle states are localized in the radial direction. Suppose the potential at the edge of the disk decays on a length scale $\xi$; then only the states localized within a strip of width $\xi$ near the edge will be significantly affected by the edge potential. In the $\mathbb{R} \times S^2$ geometry, single-particle states are localized in the azimuthal direction with $m=S_0$ corresponding to a state near the north pole and $m=-S_0$ localized near the south pole. Adding a perturbation $\psi^{\dagger}(S_z/S_0 - 1)^{\gamma}\psi$ for some large power $\gamma$ therefore corresponds to sharply increasing the energy of the states near the south pole without affecting the rest very much; such a perturbation is analogous to adding an edge potential to the disk geometry if we associate the north (south) pole of the sphere with $r=0$ ($r=R$) on the disk. In the spin language, this perturbation behaves similarly to a magnetic field.

To estimate the strength of this perturbation, we note that the electron density in the disk is $N/\pi R^2$. Suppose the edge potential decays on a length scale $\xi$; then only the states within a strip of width $\xi$ near the edge will be significantly affected by the symmetry-breaking field. Therefore, approximately $(2\pi R \xi)(N/\pi R^2) = N\xi/R$ out of the $N$ degenerate states will be affected. That is, the fraction $\xi/R$ of the degenerate states will have a marginal perturbation applied to them (roughly speaking, $k_F$ changes for these states because their $k_z$ dispersion is shifted upward in energy); we thus expect that the strength of the ``Zeeman field" to be proportional to $\xi/R$. For a thick enough wire, $\xi/R$ should be small, so most of the single-particle states remain degenerate.

After this analysis it is straightforward to understand the fate of the phase diagram upon moving to the disk geometry. Since charge is obviously still conserved, $SU(2)$-breaking marginal perturbations affect only the pseudospin sector. Therefore, despite the fact that the whole system is gapless, a gap in the pseudospin sector is enough to guarantee perturbative stability of a phase. This immediately implies that the singlet CDW and singlet SC phases are stable at both odd and even $N$. These phases should also remain distinct. Breaking $SU(2)$ symmetry does not mix CDW and SC order parameters; in fact, in the Abelian bosonization picture it is clear that the fact that $\phi_{s1},\phi_{s2}$,... are well-defined even when the external ``field" is applied is sufficient to maintain the distinctness of these phases.

The triplet CDW/SC phase at even $N$, on the other hand, is probably not strictly speaking stable to $SU(2)$-breaking perturbations. Its gaplessness originates from fluctuations of a putative spontaneous breaking of $SU(2)$ symmetry, so explicit symmetry breaking should induce a gap of order $\xi/R$. As a result, this need not be a distinct phase, but the smallness of the gap may allow a crossover to a regime where signatures of this phase remain.

\section{The Three-Dimensional Limit}
\label{sec:largeN}

Considerable work has already been done\cite{CelliSDWStrongField,FukuyamaCDWField,AbrikosovMetalStrongField,RasoltSCInHighField,Yakovenko, MiuraBiSemiconducting,HirumaFieldBi,IyeGraphite} on bulk 3D crystals in the zeroth Landau level; to compare with those results, we wish to take the bulk limit in our treatment. In the disk geometry, this means taking the radius of the wire to infinity at fixed magnetic field and carrier density. Since the Landau level degeneracy goes as the total flux penetrating the wire, the bulk limit is that of large $N$, a limit we can also take in the sphere geometry. One key expectation is that as the system becomes less one-dimensional, true long-range order appears instead of quasi-long-range order. In this section, we compare to previous work and to this expectation.

The simplest way to see the bulk limit emerge is by examining what power laws appear in correlation functions of various order parameters. Looking at our Abelian bosonization results in Section \ref{sec:phaseID} and generalizing the pattern of basis changes, we expect that the singlet order parameters obey
\begin{align}
\Delta_{CDW} = \sum_m e^{i \phi_m} \propto e^{i \phi_c/\sqrt{N}}\\
\Delta_{SC} = \sum_m e^{i \theta_m} \propto e^{i \theta_c/\sqrt{N}}
\end{align}
where $\phi_c = (\sum_m \phi_m)/\sqrt{N}$ and we have dropped the spin sector pieces of the order parameters. We saw at small $N$ that at the fixed point, the power law correlations come entirely from the $U(1)$ charge sector; the spin sector delivers constant factors. Assuming this trend continues, for a given Luttinger parameter $K$ of the charge sector, we compute that 
\begin{align}
\langle \Delta_{CDW}(x)\Delta_{CDW}^{\ast}(0) \rangle &\stackrel{SU(N)}{\sim} \frac{1}{|x|^{2/(
NK)}}\\
\langle \Delta_{SC}(x)\Delta_{SC}^{\ast}(0) \rangle &\stackrel{SO(N),USp(N)}{\sim} \frac{1}{|x|^{2K/N}}
\end{align}
Suppose that the interactions before projecting to the ZLL are fixed and weak. As $N$ grows, none of the projected interaction strengths should diverge; that is, $g_0$ should not grow with $N$. This means that corrections to the free value $K = 1$, which are controlled by the small parameter $g_0$, do not diverge with $N$. Hence as $N \rightarrow \infty$, the power law falls off slower and slower, eventually becoming a distance-independent contribution to the correlation function. This is how true long-range order appears in the bulk limit. 

It would be nice to check our conjectures about the general-$N$ phase diagram at large $N$. The starting point would be to expand the RG coefficients $\beta_{S'S''}^{S}$ at large $N$ by expanding the Wigner $6j$-symbols appearing in Eq. \eqref{eqn:explicitBetas} at large $S_0$. Unfortunately, the leading-order term in the expansion\cite{AngularMomentumBook} is proportional to the Clebsch-Gordan coefficient $\braket{S,m=0}{S',m=0;S'',m=0}$, which is precisely zero when $S+S'+S''$ is odd. If $S+S'+S''$ is even, then $\beta_{SS'}^{S''}$ is instead zero due to the selection rules in Eq. \eqref{eqn:explicitBetas}. To get any nontrivial flow, then, $1/N$ corrections must be considered, which considerably complicates the analysis.

Although it is difficult to analyze the large-$N$ limit in more detail, we can make some simple comparisons with the results of Ref. \onlinecite{Yakovenko}, where fully three-dimensional spinless fermions were considered in the parquet approximation. Ref. \onlinecite{Yakovenko} finds two zero-temperature phases in the bulk limit depending on whether the contact interactions are repulsive or attractive. In the former case, there is a transition to a CDW state, and in the latter the system is a marginal Fermi liquid. 

We do find two phases much like those above. Our CDW state exists at all $N$ and becomes long-range order in the $N \rightarrow \infty$ limit; this should be analogous to the CDW phase in Ref. \onlinecite{Yakovenko}. The marginal Fermi liquid phase is harder to compare because we have focused on $T=0$ while Ref. \onlinecite{Yakovenko} finds susceptibilities at $T>0$ which diverge only as $T\rightarrow 0$. However, the marginal Fermi liquid phase has a divergent SC susceptibility and finite CDW susceptibility as $T\rightarrow 0$, which is qualitatively similar to our $SO(N)$ ($USp(N)$) phase. 

We do find more phases than Ref. \onlinecite{Yakovenko}, in that we find a Luttinger liquid phase at all $N$ and a phase with fluctuating triplet order parameters at even $N$. A likely reason for this inconsistency is that although we require short-range interactions, we do not constrain the range of the interactions compared to the magnetic length. Ref. \onlinecite{Yakovenko} does make this assumption in order to argue that considering a projected contact interaction is sufficient, and therefore is in a special case of our results. Another possibility is that as $N$ gets large, our additional phases occupy a fraction of the phase diagram which approaches zero; we cannot rule this out because we do not know how the basin of attraction of these fixed points behaves as a function of $N$.

Beyond these considerations, it is possible for our model to break down entirely in the bulk limit due to disorder. As the wire gets thicker, it is more likely to be disordered, which would broaden the Landau levels. In fact, this would be like analyzing our pseudospin model with a random $SU(2)$-breaking field.

\section{Discussion}
\label{sec:discussion}

We first briefly summarize our main results. We mapped an interacting metallic wire with a strong magnetic field along its length to one-dimensional fermions of pseudospin $S_0 = (N-1)/2$, where $N$ is the degeneracy of the zeroth Landau level at fixed $k_x$. We then computed the phase diagram. For all $N$ and any interactions, there is spin-charge separation with a gapless charge sector (so long as the filling is incommensurate). For all $N$, there is a Luttinger liquid phase where the interactions only provide logarithmic corrections to correlations in the pseudospin part of the free theory. For $N>2$, there are also two pseudospin-gapped phases where an order parameter has power law correlations with a power that depends on $N$: a fluctuating pseudospin-singlet CDW phase and a fluctuating pseudospin-singlet SC phase. For $N$ odd, the transition between these phases is first-order, but for $N$ even, the transition is permitted to be second-order and governed by the $\mathfrak{su}(2)_{N/2}/\mathfrak{u}(1)$ parafermion CFT. Even $N>2$ has an additional phase which has no pseudospin gap and has power-law correlations of both the pseudospin-triplet CDW and SC order parameters.

Recalling that tuning $N$ is like tuning the magnetic field, our main predictions which are interesting to search for in experiments are: power law correlation functions whose power law is tuned by magnetic field, using the magnetic field to tune between a Luttinger liquid, fluctuating SC order, and CDW orders (although the extent to which this is possible depends on the details of how the interactions project at different $N$), and signatures of the phase with fluctuating pseudospin-triplet orders. One important consideration for any such experimental search is how practical the limits we are considering are for real experimental systems. The main constraint is that the carrier density must be low enough that all carriers are in the zeroth Landau level. For electrons with a quadratic dispersion, this means that the chemical potential in field must be below the energy of the first Landau level, i.e. 
\begin{equation}
\frac{\hbar^2 k_F^2}{2m} \leq \frac{\hbar eB}{m}
\label{eqn:SchrodCondition}
\end{equation}
where $m$ is the effective mass and $k_F$ is the Fermi wavevector. For a Weyl semimetal with Weyl  points at $\bv{k} = \pm k_W \hat{\bv{x}}$ (with $k_W>0$), the corresponding estimate is
\begin{equation}
\hbar |k_F-k_W| v_F \leq v_F \sqrt{\hbar eB}
\label{eqn:WeylCondition}
\end{equation}
with $v_F$ the Fermi velocity. The Landau level degeneracy $N$ in both cases is of order $\pi R^2 B/\Phi_0$, where $R$ is the wire radius and $\Phi_0 = h/e$ is the flux quantum. The LL degeneracy can be used to relate $k_F$ to the carrier density, which can then be plugged into Eqs. \eqref{eqn:SchrodCondition} and \eqref{eqn:SchrodCondition} to estimate
\begin{equation}
B \gtrsim \frac{\hbar}{e}\left(2\pi^4  n^2\right)^{1/3}
\end{equation}
for Schrodinger electrons and 
\begin{equation}
B \gtrsim \frac{\hbar}{e}\left(4\pi^4  n^2\right)^{1/3}
\end{equation}
for Weyl electrons. Assuming $n \sim 10^{17}$ cm$^{-3}$, this is about $8$ T for Schrodinger and $10$ T for Weyl. However, in both cases
\begin{equation}
N \sim 60 \left(\frac{B}{8 \text{ T}}\right)\left(\frac{R}{100\text{ nm}}\right)^2
\end{equation}
In the previous section, we saw that the power law correlation functions are most one-dimensional when $N$ is small; large $N$ quickly starts to look like long range order. Given these estimations, the large-$N$ limit should be experimentally achievable, but the small-$N$ limit may require extremely narrow wires or extremely low carrier density (to reduce the magnetic field required).

On the theoretical side, this work raises a number of open questions. Analyzing the pseudospin-gapless phase at even $N$ and its stability to $SU(2)$-breaking perturbations is an interesting and nontrivial CFT problem. Studying the various phase transitions in this model and distinguishing first-order and second-order transitions more clearly is also an interesting technical challenge in both the Abelian and non-Abelian bosonization languages. Another interesting possibility is to see if there is a deep connection with the Haldane conjecture. In particular, changing from even to odd $N$ corresponds to moving between half-integer and integer pseudospin, and the appearance of a pseudospin-gapless phase for half-integer spin is reminiscent of the Haldane conjecture. The connection is not obvious because our results are at incommensurate filling and because the set of allowed operators looks different.

\begin{acknowledgments}
We would like to thank Ian Affleck, Yingfei Gu, Pavan Hosur, Steve Kivelson, and Sri Raghu for illuminating discussions. DB is supported by the National Science Foundation under grant No. DGE-114747. CMJ's research was in part completed in Stanford University under the support of the David and Lucile Packard
Foundation. CMJ's research at KITP is supported by a fellowship from the Gordon and Betty Moore Foundation (Grant 4304). XLQ is supported by the National Science Foundation under grant No. DMR-1151786 and by the David and Lucile Packard Foundation.
\end{acknowledgments}
\appendix

\section{The Basis of Fermion Bilinears}
\label{app:bilinears}

In this appendix, we construct the matrices $M^{S,\alpha}$ with the properties discussed in Section \ref{sec:interactions} and use them to write down the $SU(2)$-invariant Hamiltonian Eq. \eqref{eqn:interactionH}, prove that the odd-$S$ matrices form a $\mathfrak{usp}(N)$ ($\mathfrak{so}(N)$) subalgebra for $N$ even (odd), and prove that the corresponding affine subalgebra of fermion bilinears has level 1 (2).

We start with some intuition. Fermion bilinears are objects $\psi^{\dagger}_mM_{mn}^{S,\alpha}\psi_n$ (suppressing the L/R indices) which transform under $SU(2)$ as $\psi^{\dagger}_{m'}U^{\dagger}_{m'm}M_{mn}^{S,\alpha}U_{nn'}\psi_{n'}$. We are thus taking two objects, one which transforms as pseudospin $S_0 = (N-1)/2$ and one which transforms as its complex conjugate, and producing an object which transforms in a pseudospin-$S$ representation. In $SU(2)$, moving from a representation to the complex conjugate is the same as time reversal. Therefore, we expect a relationship between $M_{mn}^{S,\alpha}$ and the Clebsch-Gordan coefficient $\braket{S_0,m;S_0,-n}{S,p}$ for some appropriate relationship between $p$ and $\alpha$.
Let us make this precise. 

Define a compact notation
\begin{equation}
C^{S,p}_{mn} = \braket{S_0,m;S_0,n}{S,p}
\end{equation}
for the Clebsch-Gordan coefficients fusing two spin-$S_0$ objects with $S_z$ quantum numbers $m$ and $n$ to a spin-$S$ object with $S_z$ quantum number $p$. Here $m,n = -S_0, -S_0+1...S_0$ and $p = -S, -S+1,...S$; note that $p$ and $S$ are always integers. Treating $m$ and $n$ as matrix indices, the Clebsch-Gordan coefficients are not Hermitian. Before building Hermitian matrices from them, we need to establish some preliminary properties. Using a convention where all Clebsch-Gordan coefficients are real, elementary symmetry and completeness properties of the Clebsch-Gordan coefficients lead to the identities
\begin{align}
C^{S,p}_{mn} &\propto \delta_{m+n,p} \label{eqn:CGconservation}\\
(C^{S,p})^{\dagger} &= (-1)^{S+2S_0}C^{S,p} \label{eqn:CGdagger} \\
\tr\left[(C^{S,p})^{\dagger}C^{S',p'}\right] &= \delta^{S,S'}\delta^{p,p'} \label{eqn:CGorthog}
\end{align}
In taking the Hermitian conjugate and the trace, we are treating $m$ and $n$ as the matrix indices and $S,S',p,p'$ as labels. Eq. (\ref{eqn:CGdagger}) relies on the fact that $S$ is an integer, or else there could be an extra negative sign.

Next, in the convention where the spin-$S_0$ matrices $S^x$ and $S^z$ are purely real and $S^y$ is purely imaginary, the time reversal operator is
\begin{equation}
T \equiv \Omega \mathcal{K}
\end{equation}
with $\mathcal{K}$ the antiunitary complex conjugation operator and $\Omega$ the unitary matrix $\Omega = \exp(i\pi S^y/\sqrt{2})$ (the factor of $\sqrt{2}$ is due to our normalization convention for the structure constants of $\mathfrak{su}(2)$). The matrix elements of $\Omega$ are $\Omega_{mn} = (-1)^{S_0-m}\delta_{m,-n}$; note that $\Omega^{\dagger} = (-1)^{2S_0}\Omega$ and $\Omega^2 = (-1)^{2S_0}$.

Next, define for each $p$ the matrices $A^{S,p} = C^{S,p}\Omega$. By inspection $A$ is related to the Clebsch-Gordan coefficient $C^{S,p}_{m,-n}$, as expected intuitively. Moreover, time-reversal symmetry of the $C$s implies 
\begin{equation}
C^{S,p}\Omega = (-1)^{S-p}\Omega C^{S,-p}
\label{eqn:CTimeReverse}
\end{equation}
, which can be combined with Eq. \eqref{eqn:CGdagger} and $\eqref{eqn:CGorthog}$ to find
\begin{align}
(A^{S,p})^{\dagger} &= (-1)^p A^{S,-p}\\
\tr\left(A^{S,p}A^{S',p'}\right) &= (-1)^p\delta_{S,S'}\delta_{p,-p'} \label{eqn:Aorthog}
\end{align}
Finally, we can define our desired matrices. For $\alpha =-S,-S+1,...,S$, define (suppressing matrix indices)
\begin{equation}
M^{S,\alpha} = \begin{cases}
\sqrt{\frac{k}{2}}\left(A^{S,\alpha} + (-1)^{\alpha}A^{S,-\alpha}\right) & \alpha > 0\\
\sqrt{k}A^{S,0} & \alpha = 0\\
i\sqrt{\frac{k}{2}}\left[A^{S,\alpha} - (-1)^{\alpha}A^{S,-\alpha}\right] & \alpha < 0
\end{cases}
\label{eqn:Mdefinition}
\end{equation}
Hermiticity follows immediately. Property (1) of Section \ref{sec:interactions} is satisfied by definition. Additionally using Eqs. (\ref{eqn:CGconservation}) and (\ref{eqn:CGorthog}) shows that these matrices are orthogonal and normalized according to Property (3) (all of the factors of $(-1)^p$ work out properly).

To check the transformation properties under $SU(2)$, note first that $S^z$ anticommutes with $\Omega$; this immediately proves
\begin{equation}
[S_z, A^{S,p}] = \sqrt{2}p A^{S,p} 
\label{eqn:AtransformSz}
\end{equation}
(where again the $\sqrt{2}$ is due to normalization).

Likewise, $S_x$ and $S_y$ anticommute and commute, respectively, with $\Omega$. Moreover, transforming the lower indices of a Clebsch-Gordan coefficient is the same as transforming the upper index, that is,
\begin{equation}
S^{\pm}C^{S,p} + C^{S,p}(S^{\pm})^{\dagger} = \sqrt{S(S+1)-p(p\pm 1)}C^{S,p\pm 1}
\end{equation}
These two facts imply
\begin{equation}
[S^{\pm},A^{S,p}] = \sqrt{S(S+1)-p(p\pm 1)}A^{S,p \pm 1}
\end{equation}
as desired.

From the transformation properties, it is straightforward to show that $SU(2)$ invariance requires that the interaction Hamiltonian has the form
\begin{equation}
H_{int} = \sum_{S,p} g_S (-1)^p \psi^{\dagger}_L A^{S,p}\psi_L \psi^{\dagger}_R A^{S,-p}\psi_R 
\label{eqn:interactionMomTransfer}
\end{equation}
Substituting the definition Eq. \eqref{eqn:Mdefinition} of the $M$s into Eq. \eqref{eqn:interactionH} proves that Eq. \eqref{eqn:interactionH} is the same as Eq. \eqref{eqn:interactionMomTransfer}. That is, the $M$s are just a basis rearrangement of the $A$s used to ensure Hermiticity. This is particularly clear for $S=1$; it is easy to check that $A^{1,\pm 1} \propto S^{\pm}$, so $M^{1,\pm 1} \propto S^x, S^y$ respectively. We use Eq. \eqref{eqn:interactionH} rather than Eq. \eqref{eqn:interactionMomTransfer} because the orthogonality and normalization of the $M^{S,\alpha}$ is slightly simpler than that of the $A^{S,p}$.

Having discussed the $SU(2)$ properties of the $M^{S,\alpha}$, we now demonstrate that the $M^{S,\alpha}$ for odd $S$ generate $\mathfrak{sp}(N)$ and $\mathfrak{so}(N)$ when $N$ is even and odd respectively.

It is easy to count that when $N$ is even and odd respectively, there are $N(N+1)/2$ and $N(N-1)/2$ (mutually orthogonal in the trace norm) matrices $M^{S,\alpha}$ with odd $S$; these are the dimensions of $\mathfrak{sp}(N)$ and $\mathfrak{so}(N)$ respectively. Next, note that $\Omega$ is always real and is antisymmetric (symmetric) for $N$ even (odd); therefore, we can use it as a symplectic (symmetric) form and the fundamental representation of the Lie group $USp(N)$ ($SO(N)$) consists of unitary $N \times N$ matrices $B$ which obey $B^T \Omega B=\Omega$. Passing to the Lie algebra and using $\Omega^2 = (-1)^{N+1}$, this means that if $\Omega(M^{S,\alpha})^T\Omega = (-1)^{N+1}M^{S,\alpha}$ for all odd $S$ and each $\alpha$, then $M^{S,\alpha}$ generate $\mathfrak{sp}(N)$ and $\mathfrak{so}(N)$ respectively. Using Eq. \eqref{eqn:CTimeReverse} we find
\begin{align}
\Omega A^{S,p} \Omega = (-1)^{N+1}\Omega C^{S,p} = (-1)^{S-p+N+1}A^{S,-p}
\end{align}
This immediately implies $\Omega (M^{S,\alpha})^T \Omega = (-1)^{S+N+1} M^{S,\alpha}$, which is the desired identity.

Finally, we determine the level of the $\mathfrak{so}(N)$ and $\mathfrak{usp}(N)$ affine algebras generated by the corresponding fermion bilinears. According to Eq. \eqref{eqn:bilinearOPE}, if $M^{S,\alpha}$ is any generator in the subalgebra, then the level of the corresponding affine subalgebra is $\tr(M^{S,\alpha})^2$ provided that the normalization of the subalgebra structure factors $f^{ab}_c$ is such that
\begin{equation}
\sum_{ab} f^{ab}_cf^{ab}_d = 2g\delta_{cd}
\label{eqn:structureFactorProduct}
\end{equation}
where $a,b,c,d$ label generators of the subalgebra and $g$ is the dual Coxeter number of the subalgebra. In our current normalization, $\tr(M^{S,\alpha})^2 = k$; we still need to check the normalization of the structure factors. Since the normalization Eq. \eqref{eqn:structureFactorProduct} is independent of the index $c$, we can choose the generator $c$ to be $M^{1,0} = S_z$ for convenience. From now on we will use $S',S''$ as dummy indices taking only odd values from $1$ to $N-1$ if $N$ is even and from $1$ to $N-2$ if $N$ is odd. From the definition of the structure factors it is easy to see that
\begin{align}
f^{S',\alpha;S'',\beta}_{1,0} = \frac{1}{ik}\tr\left([M^{S',\alpha},M^{S'',\beta}]M^{1,0}\right) 
\end{align}
Plugging this into Eq. \eqref{eqn:structureFactorProduct} and comparing to Eq. \eqref{eqn:MProductBeta}, we see that
\begin{equation}
\sum_{S',S'',\alpha,\beta} (f^{S',\alpha;S'',\beta}_{1,0})^2 = -\sum_{S',S''}\beta_{S',S''}^1
\end{equation}
Plugging in the definitions of the $M$s in terms of $A$s, expanding carefully and doing some reindexing turns this into
\begin{widetext}
\begin{align}
\sum_{S',S'',\alpha,\beta} (f^{S',\alpha;S'',\beta}_{1,0})^2 &= -\sum_{S',S'',\alpha \beta} (-1)^{\alpha + \beta }\tr\left(\left[A^{S',\alpha},A^{S'',\beta}\right]M^{1,0}\right)\tr \left(\left[A^{S',-\alpha},A^{S'',-\beta}\right]M^{1,0}\right)\\
&=-\sum_{S',S'',\alpha \beta} (-1)^{\alpha + \beta }\tr\left(\left[M^{1,0},A^{S',\alpha}\right]A^{S'',\beta}\right)\tr \left(\left[M^{1,0},A^{S',-\alpha}\right]A^{S'',-\beta}\right)\\
&= \sum_{S',S''\alpha \beta} 2\alpha^2 (-1)^{\alpha + \beta}\delta_{\alpha, -\beta}\delta_{S',S''} \\
&= \frac{2}{3}\sum_{S' \text{odd}}S'(S'+1)(2S'+1)\\
&= \begin{cases}
2k\left(1+\frac{N}{2}\right) & N \text{ even}\\
k(N-2) & N \text{ odd}
\end{cases}\\
&= \begin{cases}
2kg_{\mathfrak{sp}(N)} & N \text{ even}\\
kg_{\mathfrak{so}(N)} & N \text{ odd}
\end{cases}
\end{align}
\end{widetext}
where we used Eqs. \eqref{eqn:AtransformSz} and \eqref{eqn:Aorthog} to evaluate the commutators and traces. Since $\tr(M^2)= k$, we immediately read off that the level of the $\mathfrak{sp}(N)$ ($\mathfrak{so}(N)$) affine algebra is 1 (2) for $N$ even (odd). 

\section{Derivation of the RG Coefficients}
\label{app:RGderivation}

In this Appendix, we outline the derivation of Eq. \eqref{eqn:explicitBetas} starting from Eqs. \eqref{eqn:MProductBeta} and \eqref{eqn:Mdefinition}. 

The left-hand side of Eq. \eqref{eqn:MProductBeta} is $SU(2)$ invariant, so the right-hand side must be independent of $\gamma$. For convenience we sum over $\gamma$:
\begin{equation}
\beta^S_{S',S''} = \frac{1}{k^2(2S+1)}\sum_{\alpha \beta \gamma} \tr\left(\left[M^{S',\alpha},M^{S'',\beta}\right]M^{S,\gamma}\right)^2
\end{equation}

Next we plug in the explicit expression Eq. \eqref{eqn:Mdefinition} of the $M$ matrices. A careful expansion of the squares and some reindexing leads to
\begin{widetext}
\begin{align}
\beta^S_{S',S''} &= \frac{k}{(2S+1)}\sum_{\alpha \beta \gamma} (-1)^{\alpha + \beta + \gamma}\tr\left(\left[A^{S',\alpha},A^{S'',\beta}\right]A^{S,\gamma}\right)\tr\left(\left[A^{S',-\alpha},A^{S'',-\beta}\right]A^{S,-\gamma}\right)\\
&=\frac{k}{(2S+1)}\sum_{\alpha \beta \gamma} (-1)^{\alpha + \beta + \gamma}\tr\left(\left[C^{S',\alpha}\Omega,C^{S'',\beta}\Omega\right]C^{S,\gamma}\Omega\right)\tr\left(\left[C^{S',-\alpha}\Omega,C^{S'',-\beta}\Omega\right]C^{S,-\gamma}\Omega\right) \label{eqn:expandedBeta}
\end{align}
\end{widetext}
For the moment we ignore the sums on Greek indices and the commutators in order to evaluate traces of products of three Clebsch-Gordan (C-G) coefficients. Using Eq. \eqref{eqn:CTimeReverse}, we have
\begin{align}
\tr&\left(C^{S',\alpha}\Omega C^{S'',\beta}\Omega C^{S,\gamma}\Omega\right) \nonumber\\ &=(-1)^{S''-\beta+2S_0}\tr\left(C^{S',\alpha} C^{S'',-\beta} C^{S,\gamma}\Omega\right)\\
&= (-1)^{S''-\beta+2S_0}\sum_{mnl}(-1)^{S_0+m}C^{S',\alpha}_{mn} C^{S'',-\beta}_{nl} C^{S,\gamma}_{l,-m}
\end{align}
Note that this is only nonzero when $m+n=\alpha$, $n+l=-\beta$, and $l-m = \gamma$, which means $\alpha+\beta+\gamma=0$. This removes a phase factor in Eq. \eqref{eqn:expandedBeta}. Transposing the first term using Eq. \eqref{eqn:CGdagger} manipulates this equation into a form for which there is a known\cite{AngularMomentumBook} identity relating such a product of three C-G coefficients to a product of a 6j symbol and another C-G coefficient. Applying the identity, we get
\begin{align}
\tr&\left(C^{S',\alpha}\Omega C^{S'',\beta}\Omega C^{S,\gamma}\Omega\right) \nonumber\\
&= (-1)^{S'+S-\beta}\sqrt{(2S+1)(2S'+1)}\tilde{C}^{S''-\beta}_{S',\alpha;S,\gamma} \begin{Bmatrix}
S_0 & S_0 & S'\\
S'' & S & S_0
\end{Bmatrix}
\end{align}
where $\tilde{C}$ is a C-G coefficient for combining spin $S$ and $S'$ into $S''$. Substituting this relationship into Eq. \eqref{eqn:expandedBeta} and using the symmetry properties of the $6j$ symbols converts it to
\begin{widetext}
\begin{align}
\beta^S_{S',S''} = k \left(\begin{Bmatrix}
S & S' & S''\\
S_0 & S_0 & S_0
\end{Bmatrix}\right)^2 \sum_{\alpha \beta \gamma} &\left((-1)^{S'-\beta}\sqrt{2S'+1}\tilde{C}^{S'',-\beta}_{S',\alpha;S,\gamma} - (-1)^{S''-\alpha}\sqrt{2S''+1}\tilde{C}^{S',-\alpha}_{S'',\beta;S,\gamma}\right) \times \nonumber \\
&\left((-1)^{S'+\beta}\sqrt{2S'+1}\tilde{C}^{S'',\beta}_{S',-\alpha;S,-\gamma} - (-1)^{S''+\alpha}\sqrt{2S''+1}\tilde{C}^{S',\alpha}_{S'',-\beta;S,-\gamma}\right)
\end{align}
\end{widetext}

Using elementary symmetry properties of the C-G coefficients, all the $\alpha$s and $\gamma$s can be placed on the bottom and given the same sign up to some phase factors and factors of $\sqrt{2S'+1}$ or $\sqrt{2S''+1}$. This allows the use of the completeness relations of the C-G coefficients in order to perform the sums over $\alpha$ and $\gamma$ and to remove all the C-G coefficients. The remaining $\beta$ dependence disappears, allowing the sum over $\beta$ to be replaced by a factor of $(2S''+1)$. These manipulations are simple but tedious; tracking all the factors carefully (and remembering that $\alpha, \beta, \gamma, S, S',$ and $S''$ are integers) produces Eq. \eqref{eqn:explicitBetas}.

\section{Selection Rules for OPEs}
\label{app:selection}

We found that in Eq. \eqref{eqn:explicitBetas} that $\beta^{S}_{S'S''} = 0$ if $S+S'+S''$ is even. In this section, we will use Young tableaux to demonstrate how this selection rule results from the symmetry properties of the fermion bilinears.

Consider the products of three $M$s as they appear in Eq. \eqref{eqn:MProductBeta}. The object $\tr(M^{S',\alpha}M^{S'',\gamma}M^{S,\delta})$ intuitively takes a spin-$S'$ and spin-$S''$ object, fuses them, and finds its overlap with the spin-$S$ channel. There are of course constraints on $\alpha$, $\gamma$, and $\delta$, but for the moment we only care about whether $\beta_{S'S''}^S$ is zero.

The symmetry of such fusions can be encoded in Young tableaux. For example, consider $S'=2$,$S''=1$. Then the two terms in the commutator $\tr([M^{2,\alpha},M^{1,\gamma}]M^{S,\delta})$ are
\begin{widetext}
\begin{align}
\ytableausetup{nosmalltableaux}
\ydiagram{4} \bigotimes \ydiagram[*(gray)]{2}
 &= \ydiagram[*(white)]{4}*[*(gray)]{6} \bigoplus \ydiagram[*(white)]{4}*[*(gray)]{5,1} \bigoplus \ydiagram[*(white)]{4}*[*(gray)]{4,2} \label{eqn:youngExample}\\
 \ytableausetup{nosmalltableaux}
\ydiagram[*(gray)]{2} \bigotimes \ydiagram{4} 
 &= \ydiagram[*(gray)]{2}*[*(white)]{6} \bigoplus \ydiagram[*(gray)]{2}*[*(white)]{5,1} \bigoplus \ydiagram[*(gray)]{2}*[*(white)]{4,2}
\label{eqn:youngExampleSwitch}
\end{align}
\end{widetext}
The shading tracks whether the box came from the spin-$2$ or the spin-$1$ representation. It is implied that all boxes with the same shading are \textit{symmetrized}, regardless of the row, because they are symmetrized on the left-hand side of Eq. \eqref{eqn:youngExample}.  The three terms correspond to $S = 3,2,1$ respectively.

It is now clear from the symmetry properties of the Young tableaux (that is, rows are symmetrized and columns are antisymmetrized) that in subtracting Eq. \eqref{eqn:youngExampleSwitch} from Eq. \eqref{eqn:youngExample} the spin-3 and spin-1 tableaux will cancel out, while the spin-2 tableau will not. The commutator in Eq. \eqref{eqn:explicitBetas} is exactly such a difference, so the commutator must produce zero if $S \neq 2$.

More generally, there will be a fully symmetric tableau with $2S'$ boxes  (the white boxes in Eq. \eqref{eqn:youngExample}) fused with a fully symmetric tableau with $2S''$ boxes (the shaded boxes in Eq. \eqref{eqn:youngExample}). Consider the fusion to spin $S$. There are $2(S'+S'')$ boxes total, $2S$ of which must be ``dangling" in the first row. Hence there are $S'+S''-S$ columns which have two boxes in them (this must be nonnegative for that fusion channel to be allowed at all), one of which must come from $S'$ and the other of which must come from $S''$. Therefore under exchange of the $S'$ and $S''$ tableaux, the wavefunction picks up a factor of $(-1)^{S'+S''-S} = (-1)^{S+S'+S''}$ (since $S$ is an integer). If $S+S'+S''$ is even, then the wavefunction is symmetric under this exchange and the commutator produces zero, so $\beta_{S'S''}^{S} = 0$. 

\section{Mean Field Theory}
\label{app:meanfield}

In this section, we explain our mean-field procedure that is used for intuition about the phase diagram. In particular, we will compute the susceptibility for each possible CDW or SC order parameter to show that at mean-field level, the most negative coupling constant produces the strongest tendency towards order (the strongest divergence in the susceptibility).

The action is
\begin{equation}
S = \int dx d\tau \sum_m \psi^{\dagger}_m \partial_{\tau} \psi_m + H_0 +H_{int}
\end{equation}
with $H_0$ defined in Eq. \eqref{eqn:H0}. We choose to write $H_{int}$ in the exchange channel as in Eq. \eqref{eqn:exchangeCoupling}.

Next, consider the fat unity
\begin{widetext}
\begin{equation}
1 \propto \int DG^{S\alpha} \exp\left(-\frac{1}{4|g_S^{ex}|}\int dx d\tau \left(G^{S,\alpha}+2 g^{E}_S \psi^{\dagger}_LM^{S, \alpha}\psi_R\right)\left(\left(G^{S,\alpha}\right)^{\ast}+2 g_S^{E} \psi^{\dagger}_RM^{S, \alpha}\psi_L\right)\right)
\end{equation}
\end{widetext}
where $G^{S,\alpha}$ is a complex bosonic field and spacetime dependences have been suppressed. Then it is easy to check by expanding that when $g_S < 0$, the quartic term produces the correct sign to cancel off the interaction. We expect no low-energy instabilities when $g_S^{ex} > 0$, so the mean field does not need to make sense. 

Defining the object $\Psi^{\dagger}(x) = \begin{pmatrix}
\psi^{\dagger}_L(x) & \psi^{\dagger}_R(x) 
\end{pmatrix}$ (a $2N$-component object) and inserting the fat unity into the path integral, the effective action is then
\begin{widetext}
\begin{equation}
S_{eff} = \int dz d\tau \left[\frac{1}{4|g_S^{ex}|}|G^{S,\alpha}|^2 + \Psi^{\dagger} \begin{pmatrix}
G_{0,L}^{-1} & -\frac{1}{2} G^{S,\alpha} M^{S,\alpha}\\
-\frac{1}{2} (G^{S,\alpha})^{\ast} M^{S,\alpha} & G_{0,R}^{-1}
\end{pmatrix} \Psi \right] \label{eqn:exchangeSeff}
\end{equation}
\end{widetext}
where $G_{0,L(R)}$ is the noninteracting Green's function for the left (right) movers (and is independent of $m$). We now integrate out the fermions and expand to second order in $G^{S,\alpha}$. The expansion produces terms in the free energy proportional to $\tr(M^{S,\alpha}M^{S',\beta})G^{S,\alpha}(G^{S',\beta})^{\ast}$; thanks to our convenient choice of the $M$s, the trace collapses the sum to only the diagonal terms. Hence at second order, all the order parameters decouple, yielding the free energy
\begin{equation}
F \approx  \int dq d\omega \frac{|G^{S,\alpha}(q,\omega)|^2}{4|g_S^{ex}|}\left[1 + \chi_{CDW}(q,\omega)\right] \label{eqn:exchangeF}
\end{equation}
The linear term vanishes by the trace in L/R space, and we have dropped the zeroth-order (free fermion) contribution. We have defined the CDW susceptibility
\begin{equation}
\chi_{CDW}(q,\omega) = k|g_S^{ex}|\sum_{p,\omega'}G_{0,L}(p,\omega')G_{0,R}(p-q,\omega'-\omega)
\end{equation}
The trace over the flavor index produces the factor of $k$. Here $\omega$ and $\omega'$ are bosonic Matsubara frequencies. We have assumed that all $g_S^{E} < 0$, and there are implicit sums over all $S,\alpha$.

Evaluating the sum of noninteracting fermionic Green functions by standard techniques produces, at zero temperature and zero frequency, the static susceptibility
\begin{equation}
\chi_{CDW}(q,\omega = 0) = k |g^E_S|\pi \log \left| \frac{(\delta q)^2}{4\Lambda^2 - (\delta q)^2} \frac{4k_F - 2\Lambda - \delta q}{4k_F + 2\Lambda - \delta q}\right|
\end{equation}
Here $\delta q = q + 2k_F$ and $\Lambda$ is the momentum cutoff of the low-energy non-interacting theory. There is a divergence at $\delta q = 0$ (i.e. $q=2k_F$) which scales as $2\pi k |g^E_S| \log \delta q$.

A completely analogous computation in the Cooper channel yields a static susceptibility
\begin{equation}
\chi_{SC}(q,\omega = 0) = k |g_S^C|\pi \log \left| \frac{q^2(4k_F+q-2\Lambda)}{(4\Lambda^2-q^2)(4k_F+q+2\Lambda)}\right|
\end{equation}
This has a $q=0$ divergence scaling as $2\pi k |g^C_S|\log q$.

The conclusion of all of this is that at mean-field level, any negative coupling constant produces a logarithmically divergent susceptibility in its corresponding channel. Moreover, the strength of the divergence is the coupling constant times a channel- and $S-$independent factor. Therefore, all of the coupling constants are directly comparable, and the most negative coupling constant should produce the strongest tendency towards order. 

Since there is no spontaneous symmetry breaking of a continuous symmetry in one dimension, we expect that there are significant corrections to the mean field picture. First, decoupling of the order parameters should not persist past second order, Second, we expect long-range, mean-field order to be corrected to quasi-long-range order. As a heuristic guide, then, we expect that the channel with the most negative coupling constant will have quasi-long-range order and that other channels will not.

\bibstyle{apsrev4-1} \bibliography{references}

\end{document}